\documentclass{aastex62}
\usepackage{CJK}
\usepackage{amsmath}
\graphicspath{{./}{figures/}}

\received{}
\revised{}
\accepted{}
\submitjournal{ApJ}

\shorttitle{Alfv\'en waves}
\shortauthors{Shi et al.}

\begin{document}
\begin{CJK*}{UTF8}{gbsn}
\title{Propagation of Alfv\'en waves in the expanding solar wind with the fast-slow stream interaction}

\correspondingauthor{Chen Shi}
\email{cshi1993@g.ucla.edu}

\author[0000-0002-2582-7085]{Chen Shi (时辰)}
\affiliation{Earth,Planetary, and Space Sciences, University of California, Los Angeles \\
    Los Angeles, CA 90095, USA}
    
\author{Marco Velli}
\affiliation{Earth,Planetary, and Space Sciences, University of California, Los Angeles \\
    Los Angeles, CA 90095, USA}

\author{Anna Tenerani}
\affiliation{Department of Physics, The University of Texas at Austin, \\
     TX 78712, USA}

\author{Franco Rappazzo}
\affiliation{Earth,Planetary, and Space Sciences, University of California, Los Angeles \\
    Los Angeles, CA 90095, USA}

\author{Victor R\'eville}
\affiliation{IRAP, Universit\'e Toulouse III - Paul Sabatier,
CNRS, CNES, Toulouse, France}

\begin{abstract}
We carry out two-dimensional magnetohydrodynamic (MHD) simulations of an ensemble of Alfv\'enic fluctuations propagating in a structured, expanding solar wind including the presence of fast and slow solar wind streams. Using an appropriate expanding box model, the simulations incorporate the effects of fast-slow stream shear and compression and rarefaction self-consistently. We investigate the radial and longitudinal evolution of the cross helicity, the total and residual energies and the power spectra of outward and inward Alfv\'enic fluctuations. The stream interaction is found to strongly affect the radial evolution of Alfv\'enic turbulence. The total energy in the Alfv\'en waves is depleted within the velocity shear regions, accompanied by the decrease of the normalized cross helicity. The presence of stream-compression facilitates this process. Residual energy fluctuates around zero due to the correlation and de-correlation between the inward/outward waves but no net growth or decrease of the residual energy is observed. The radial power spectra of the inward/outward Alfv\'en waves show significant longitudinal variations. Kolmogorov-like spectra are developed only inside the fast and slow streams and when both the compression and shear are present. On the other hand, the spectra along the longitudinal direction show clear Kolmogorov-like inertial ranges in all cases.
\end{abstract}



\section{Introduction}\label{sec:intro}
Turbulence is one of the most important phenomena in space. Inside the heliosphere, it is believed to be fundamental to various physical processes such as the heating and acceleration of the solar corona and wind and the acceleration and propagation of energetic particles, etc. Thus, understanding solar wind turbulence has been one of the most crucial tasks in heliospheric physics and astrophysics as a whole. In addition, as direct measurements of solar wind accumulate, the wind itself serves as a laboratory for the study of the fundamental physics of plasma turbulence.

The study of fluctuations in the solar wind began in the 1960s, when \citet{coleman1968}, using Mariner 2 data, created the first frequency spectra of magnetic field energy and showed that they were power-laws compatible with the well known Kolmogorov power-law spectrum.
He therefore  proposed that this turbulence was created by the differential flow in fast and slow solar wind streams. \citet{belcherdavis1971}, by analyzing Mariner 5 data, showed however that much of the fluctuations comprising the turbulence, especially those in the fast streams, have correlations between velocity and magnetic field consistent with large amplitude Alfv\'en waves propagating away from the sun, and should therefore be generated somewhere close to the sun and propagate outward. A sequence of theoretical works on the propagation of Alfv\'en waves were developed, starting from linear models, both WKB theory \citep[e.g.][]{alazraki1971,belcher1971,hollweg1974} and non-WKB theory \citep[e.g.][]{heinemann1980}. However, whether the Alfv\'en waves are generated near the sun or evolve dynamically in the solar wind was still a problem. In fact one major progress in the theory of incompressible MHD turbulence, namely dynamical alignment \citep[e.g.][]{dobrowolny1980}, was developed to understand the dominance of outwardly propagating Alfv\'enic fluctuations in the solar wind as a result of an ongoing nonlinear cascade. 

In the 1980s, with the Helios data, it was found that the energy spectra of the turbulence steepen toward the well-known Kolmogrov's $5/3$ law or Iroshnikov-Kraichnan's $3/2$ law \citep{iroshnikov1964,kraichnan1965}, indicating nonlinear processes in the evolution of the fluctuations \citep{bavassano1982}. Meanwhile, MHD turbulence transport models were developed. For example, \citet{tuetal1984} established a WKB-like Alfv\'enic turbulence model which reproduces the steepening of the energy spectrum successfully. However, in this model the energy of inward propagating Alfv\'enic fluctuations, required for a nonlinear cascade, is an input parameter rather than the self-consistent outcome of the calculation. Thus understanding the generation and the propagation of the inward propagating Alfv\'en waves is crucial. \citet{roberts1987a,roberts1987b} analyzed the Voyager and Helios data and concluded that the dominance of the outward propagation weakens as the heliocentric distance increases. They also proposed that the outward propagating fluctuations are generated near the sun but the sheared streams in the solar wind accounts for the generation of the inward waves. \citet{grappinandmangeney1990} and \citet{marsch1990} studied the radial evolution and the properties of Alfv\'enic turbulence spectra and their dependence on various solar wind parameters by analyzing the Helios data by means of Els\"asser variables. \citet{roberts1992} carried out 2D incompressible MHD simulations with large-scale velocity shear and isotropic Alfv\'enic fluctuations. Their results showed that the velocity shear layer can produce a nonlinear cascade toward smaller scale fluctuations and the normalized cross helicity, i.e. the relative abundance of the outward propagating Alfv\'en waves, is decreased systematically at all wave numbers by the shear while the kinetic and magnetic energies remain approximately in equipartition \citep[see also][]{goldstein1989}. In addition to the sheared streams, the inhomogeneity of the large-scale solar wind structures due to the expansion of the solar wind leads to reflection of the outward propagating Alfv\'en waves and couples the outward and the inward waves linearly which may account for the decrease of the ``Alfv\'enicity'' \citep[e.g.][]{velli1991,velli1993}.

Although analytical modeling of MHD turbulence has been developed significantly since the 1980s \citep[e.g.][]{zhou1990,zank1996,zank2012}, limitations still exist in the models. First, various approximations must be made to close the moment equations. Whether the closures are physically correct is important and is often debatable. Second, all of the existing models deal with incompressible or nearly-incompressible turbulence while compressible effects may be important in the solar wind, especially in the fast-slow stream interaction regions. Third, the models are based on the two-scale separation method and thus the large-scale structure is not evolved self-consistently. In addition, the source terms generating the turbulence, e.g. the terms related with the velocity shear and the compressional effects, are only phenomenologically derived. Considering the drawbacks of models, it is good to adopt the direct numerical simulations (DNS) as a method to study the turbulence since the simulation solves the physical system self-consistently and can be fully compressible. However, vast computational capacity is required in order to fully simulate the turbulence evolution inside the heliosphere due to the huge separation of spatial scales. Compromising methods were developed, e.g. the expanding box model (EBM) \citep{grappin1996,tenerani2017}, which by tracking a box co-moving with the radial mean flow, neglecting the high-order curvature terms, allows one to simulate the nonlinear evolution of the waves and turbulence and the stream structures with the expansion effect taken into consideration.

In this work, we carry out 2D MHD simulations based on the expanding box model to study the propagation of Alfv\'en waves and the evolution of the turbulence in the inner heliosphere. Especially, we focus on the effects of the evolving fast-slow stream interaction present in the simulations. The simulation parameters are chosen to be close to the real solar wind conditions. We inspect the radial evolution and the longitudinal variation of some parameters that are important in the MHD turbulence study, i.e. the energy in the Els\"asser variables, the normalized cross helicity and the normalized residual energy. We show that all of the parameters are significantly affected by the velocity shear and the compression between the streams. We also investigate the power spectra of the Els\"asser variables. The paper is organized as follows. In Section \ref{sec:numerical}, we describe the numerical method that is used in this study and the setup of the simulations. In Section \ref{sec:results} we present the simulation results. In Section \ref{sec:conclusion} we conclude and discuss prospective future works.

\section{Numerical method}\label{sec:numerical}
In this section we describe the numerical method, i.e. the corotating expanding box model, in Section \ref{subsec:EBM} \& \ref{subsec:cEBM} and then present the initial setup and the choice of parameters in Section \ref{subsec:setup_sim_parameters}.
\subsection{Expanding Box Model in Conservation Form}\label{subsec:EBM}

The derivation of the expanding box model based on the convective form of the MHD equation is well described in previous papers \citep[e.g.][]{grappin1993,grappin1996}. The idea is to break the velocity $\mathbf{U}$ into two parts: the radial mean flow and the velocity in the frame of the mean flow:
\begin{equation}
    \mathbf{U} = U_0 \hat{e}_r + \mathbf{u}
\end{equation}
where $U_0$ is the constant radial speed. Simulation domain is a thin box (small radial extent) co-moving with the mean flow and the (normalized) expanding coordinate system ($\Tilde{x},\Tilde{y},\Tilde{z}$) transforms from the inertial coordinates by
\begin{equation}
    \Tilde{x} = x - R(t), \, \Tilde{y} =\frac{R_0}{R(t)} y ,  \, \Tilde{z} =\frac{R_0}{R(t)} z
\end{equation}
where $R(t) = R_0 + U_0 t$ such that the derivatives are
\begin{equation}
    \frac{\partial}{\partial x} = \frac{\partial }{\partial \Tilde{x}}, \, \frac{\partial}{\partial y} = \frac{R_0}{R(t)} \frac{\partial}{\partial \Tilde{y}}, \, \frac{\partial}{\partial z} = \frac{R_0}{R(t)} \frac{\partial}{\partial \Tilde{z}}
\end{equation}
Inside the expanding box, the mean flow can be written, by neglecting the high-order curvature terms, as
\begin{equation}\label{eq:Ur_in_cart}
    U_0 \hat{e}_r \approx U_0 \left[\hat{e}_x  + \frac{y}{R(t)} \hat{e}_y + \frac{z}{R(t)} \hat{e}_z \right]
\end{equation}
Plugging Eq (\ref{eq:Ur_in_cart}) into the MHD equation gives the EBM equation set
\begin{subequations}\label{eq:EBM_convection}
\begin{equation}\label{eq:EBM_convection_rho}
    \frac{\partial \rho}{\partial t} = -\nabla \cdot \left( \rho \mathbf{u}\right)-\frac{2}{\tau} \rho
\end{equation}
\begin{equation}\label{eq:EBM_convection_u}
   \frac{\partial \mathbf{u}}{\partial t}  =-\mathbf{u} \cdot \nabla \mathbf{u} - \frac{1}{\rho}\nabla \left(p+ \frac{1}{2}B^2 \right) + \frac{1}{\rho} \mathbf{B} \cdot \nabla \mathbf{B} - \frac{1}{\tau} \left(
   \begin{array}{ccc}
        0&0&0  \\
        0&1&0 \\
        0&0&1
   \end{array}\right) \mathbf{u}
\end{equation}
\begin{equation}
     \frac{\partial \mathbf{B}}{\partial t} = \nabla \times \left( \mathbf{u} \times \mathbf{B}  \right) - \frac{1}{\tau} \left( \begin{array}{ccc}
          2 & 0 & 0  \\
          0 & 1 & 0  \\
          0 & 0 & 1
     \end{array}\right) \mathbf{B}
\end{equation}
\begin{equation}
    \frac{\partial p}{\partial t} = - \mathbf{u} \cdot \nabla p - \kappa \left( \nabla \cdot \mathbf{u}\right) p - \frac{2\kappa}{\tau} p
\end{equation}
\end{subequations}
where $\kappa$ is the adiabatic index and $\tau = R(t) / U_0$ is the ``expansion time''. Eq (\ref{eq:EBM_convection}) is very similar to the normal MHD equation set except for: (1) The velocity field is in the reference frame of the radial mean flow. (2) New terms with the expansion time $\tau$ are introduced by the radial mean flow and they represent the expansion effect. A more detailed discussion of the EBM properties can be found in \citep{grappin1996}.

For the conservation-form of the MHD equation, care must be taken on the expansion terms. The expansion terms for the density and magnetic field equations remain unchanged as in Eq (\ref{eq:EBM_convection}) but not for the momentum and energy equations. Take the momentum equation as an example. Because the left-hand-side of the momentum equation can be written as
\begin{equation}
    \frac{\partial \left(\rho \mathbf{u}\right)}{\partial t} + \nabla \cdot \left( \rho \mathbf{uu}\right) = \left[ \rho \left( \frac{\partial \mathbf{u}}{\partial t} + \mathbf{u} \cdot \nabla \mathbf{u}\right) \right] + \left[ \frac{\partial \rho}{\partial t} + \nabla \cdot \left( \rho \mathbf{u}\right)\right] \mathbf{u}
\end{equation}
the expansion term thus consists of the part that comes from the velocity equation (Eq (\ref{eq:EBM_convection_u})) and that from the density equation (Eq (\ref{eq:EBM_convection_rho})):
\begin{equation}
    E_{m} = - \frac{1}{\tau} \left(
   \begin{array}{ccc}
        0&0&0  \\
        0&1&0 \\
        0&0&1
   \end{array}\right) \rho\mathbf{u} - \frac{2}{\tau} \rho \mathbf{u} = - \frac{1}{\tau} \left(
   \begin{array}{ccc}
        2&0&0  \\
        0&3&0 \\
        0&0&3
   \end{array}\right) \rho\mathbf{u} 
\end{equation}
Similarly, one can show that the expansion term of the energy equation
\begin{equation}
        \frac{\partial e}{\partial t} = - \nabla \cdot \left[ \left(e + p + \frac{1}{2} B^2  \right) \mathbf{u} - \left( \mathbf{u} \cdot \mathbf{B} \right) \mathbf{B} \right]
\end{equation}
where $e = \frac{p}{\kappa -1 } + \frac{1}{2} \rho u^2  + \frac{1}{2} B^2 $, is 
\begin{equation}
    E_e = - 2\frac{\kappa }{\kappa -1 } \frac{p}{\tau} - \frac{\rho}{\tau} \left( u_x^2 + 2u_y^2 + 2u_z^2 \right) - \frac{1}{\tau} \left( 2B_x^2 + B_y^2 + B_z^2 \right) 
\end{equation}
In summary, \textit{the EBM equation set in conservation form} is
\begin{subequations}
    \begin{equation}
        \frac{\partial \rho}{\partial t} = - \nabla \cdot \left(  \rho \mathbf{u} \right) -\frac{2}{\tau} \rho
    \end{equation}
    
    \begin{equation}
        \frac{\partial \left(\rho \mathbf{u} \right)}{\partial t} = - \nabla \cdot \left[\rho \mathbf{uu} + \left( p + \frac{1}{2} B^2  \right) \mathbf{I} - \mathbf{BB}  \right]  - \frac{1}{\tau} \left(
   \begin{array}{ccc}
        2&0&0  \\
        0&3&0 \\
        0&0&3
   \end{array}\right) \rho\mathbf{u} 
    \end{equation}

    \begin{equation}
        \frac{\partial \mathbf{B}}{\partial t} = \nabla \times \left( \mathbf{u} \times \mathbf{B}  \right) - \frac{1}{\tau} \left( \begin{array}{ccc}
          2 & 0 & 0  \\
          0 & 1 & 0  \\
          0 & 0 & 1
     \end{array}\right) \mathbf{B} 
    \end{equation}
    
    \begin{equation}
        \frac{\partial e}{\partial t} = - \nabla \cdot \left[ \left(e + p + \frac{1}{2} B^2  \right) \mathbf{u} - \left( \mathbf{u} \cdot \mathbf{B} \right) \mathbf{B} \right]- \frac{1}{\tau} \left[ \frac{2\kappa }{\kappa -1 } p + \rho \left( u_x^2 + 2u_y^2 + 2u_z^2 \right) + \left( 2B_x^2 + B_y^2 + B_z^2 \right) \right]
    \end{equation}
\end{subequations}
with 
\begin{equation}
    e = \frac{p}{\kappa -1 } + \frac{1}{2} \rho u^2  + \frac{1}{2} B^2 
\end{equation}

\subsection{Corotating Expanding Box}\label{subsec:cEBM}
As explained by \citep{grappin1996}, in order to simulate the compression between fast and slow streams, we need to rotate the expanding box coordinates by a small angle $\alpha$ such that the new coordinate system $\mathbf{x^\prime}$ is
\begin{equation}
    \left( \begin{array}{c}
        x^\prime \\
        y^\prime \\
        z^\prime
    \end{array} \right) = \left( 
        \begin{array}{ccc}
            \cos \alpha &  -\sin \alpha & 0\\
            \sin \alpha & \cos \alpha & 0 \\
            0 & 0 & 1 
        \end{array}\right)
        \left( \begin{array}{c}
            \tilde{x} \\
            \tilde{y} \\
            \tilde{z}
        \end{array}
            \right)
\end{equation}
The angle $\alpha$ is constant and is the initial inclination of the interface between the fast and slow streams with respect to the radial direction. The initial condition for the stream structure is
\begin{equation}
    \mathbf{u_0} = u_0(y^\prime) \hat{e}_{\tilde{x}}, \, \rho_0 = \rho_0(y^\prime)
\end{equation}
i.e. the velocity is along the radial direction but varies with $y^\prime$ instead of $y$ so that compression is induced. The temperature of the stream $T_0 = T_0(y^\prime)$ such that $p_0 = \rho_0 T_0$ is uniform.

\begin{figure}[ht!]
\centering
\includegraphics[scale=0.6]{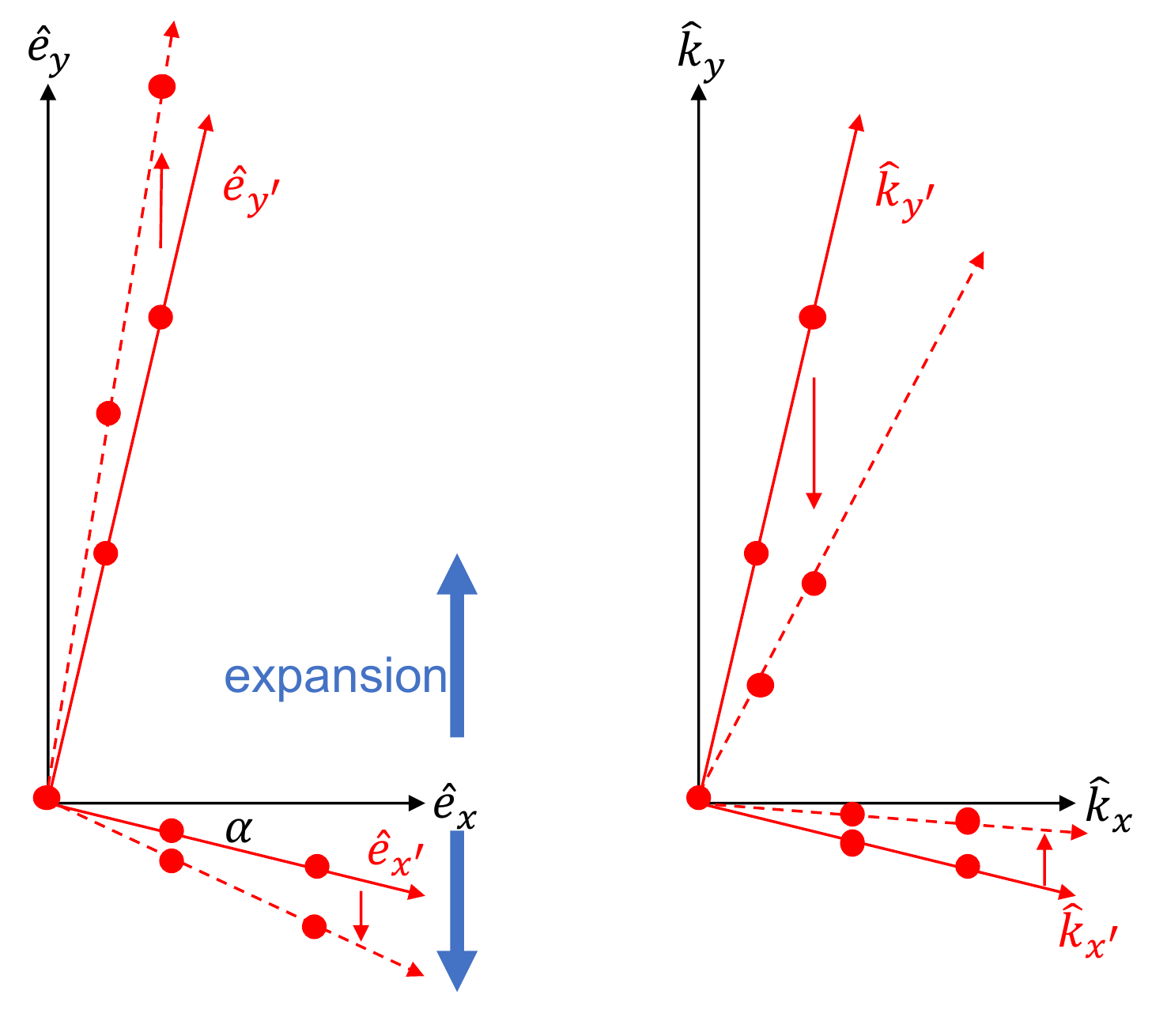}
\caption{Illustration of the deformation of the corotating expanding box coordinates $\mathbf{x^\prime}$ (left) and the wave vector $\mathbf{k_{x^\prime}}$ (right). Black axes are the normal expanding box coordinates (wave vectors) with $\hat{e}_x$ ($\hat{k}_x$) to be radial. The solid red axes represent the initial state of the corotating coordinates $\mathbf{x^\prime}$ (wave vectors $\mathbf{k_{x^\prime}}$) which is orthogonal. The red dots represent mesh gridpoints in the simulation domain. The dashed red lines are axes $\hat{e}_{x^\prime}$ ($\hat{k}_{x^\prime}$) and $\hat{e}_{y^\prime}$ ($\hat{k}_{y^\prime}$) after the simulation starts. \label{fig:illust_cEBM}}
\end{figure}

We should point out that, although the coordinates $\mathbf{x^\prime}$ are orthogonal at the beginning, they do not remain orthogonal as the box expands unless $\alpha=0$, as illustrated in the left panel of Figure \ref{fig:illust_cEBM}. The black axes show the normal expanding box coordinates with $\hat{e}_x$ aligned with the radial direction and $\hat{e}_y$ along the azimuthal ($\varphi$) direction. The solid red axes represent the initial state of the corotating expanding box coordinates $\mathbf{x^\prime}$, an orthogonal coordinate system rotated by an angle $\alpha$ with respect to the radial direction. The red dots represent a few mesh points in the simulation domain. Due to the expansion along the $\varphi$ direction, both the $\hat{e}_{x^\prime}$ and $\hat{e}_{y^\prime}$ turn away from the radial axis, as shown by the dashed red axes. That is to say, the angle between $\hat{e}_{x^\prime}$ and $\hat{e}_{y^\prime}$ becomes larger than $\pi/2$ after the simulation starts. A positive aspect of this frame is that, if we set the initial magnetic field to be aligned with $\hat{e}_{x^\prime}$
\begin{equation}\label{eq:B0_abstract}
    \mathbf{B_0} = B_0(y^\prime) \hat{e}_{x^\prime}
\end{equation}
it will remain aligned with $\hat{e}_{x^\prime}$ for all time. Thus, in all the simulations we set up $\mathbf{B_0}$ as like Eq (\ref{eq:B0_abstract}) and we will call $\hat{e}_{x^\prime}$ the parallel direction hereinafter. Note that, although the axes in real space are turned away from the radial direction, the wave vectors are actually turned toward the radial direction (right panel of Figure \ref{fig:illust_cEBM}) due to the increase of the grid spacing in $y$.

The code operates mainly in the Fourier space $(k_{x^\prime}, k_{y^\prime})$. A third-order Runge-Kutta method is used for time integration. Vectors remain defined in the $(\hat{e}_{\tilde{x}},\hat{e}_{\tilde{y}},\hat{e}_{\tilde{z}})$ directions although the mesh grid is on $(x^\prime, y^\prime)$. At each time step, fluxes are calculated in real space first and then Fourier transformed. Time advance is done in Fourier space and we need the following projection in order to transform the derivatives on $\left( x^\prime, \, y^\prime \right)$ to the derivatives on  $\left( x, \, y \right)$:
\begin{equation}
    \begin{aligned}
        &k_x = k_{x^\prime} \cos \alpha +  k_{y^\prime} \sin \alpha \\
        &k_y = \frac{R_0}{R(t)} \left(- k_{x^\prime} \sin \alpha +  k_{y^\prime} \cos \alpha \right)
    \end{aligned}
\end{equation}
$\nabla \cdot \mathbf{B} = 0$ is automatically preserved by this algorithm. Because we are interested in the evolution of turbulence, rather than heating or plasma thermodynamics,  we apply a smooth numerical filter to all fields to ensure proper de-aliasing rather than explicit viscosity or resistivity. The filter is defined in Fourier space:
\begin{equation}
     \mathbf{\hat{f}}(k_{x^\prime}, k_{y^\prime})  = \mathbf{f}(k_{x^\prime}, k_{y^\prime}) \times T(k_{x^\prime}) \times T(k_{y^\prime}) 
\end{equation}
where $\mathbf{f}$ is the field before filtering and $\mathbf{\hat{f}}$ is the field after filtering. The function $T(k)$ is the same as the fourth-order filter of the compact finite difference scheme (Eq (C.2.2) and (C.2.4) of \citep{lele1992}) with constraints $\beta = d = 0$
\begin{equation}
    T(k) = \frac{a+b\cos(w) + c \cos(2 w)}{1 + 2 \lambda \cos(w)}
\end{equation}
where $w = 2 \pi k \Delta \in \left[ -\pi, \, \pi \right]$ is the normalized wave number ($\Delta$ is the grid spacing) and $a=(5+6\lambda)/8$, $b=(1+2\lambda)/2$, $c=-(1-2\lambda)/8$ with $\lambda$ to be a free parameter in the range $[-0.5,0.5]$ (refer to Fig. 19 of \citep{lele1992} for the shape of $T(k)$). $\lambda = 0.5$ corresponds to no filtering at all. In our simulations we set $\lambda = 0.45$ such that the numerical stability is ensured without too much numerical dissipation.

\subsection{Initial Setup and Parameters}\label{subsec:setup_sim_parameters}
The initial condition consists of the large scale stream structure and the Alfv\'en waves. As mentioned in Section \ref{subsec:cEBM}, the stream structure is of the form
\begin{equation}
    \mathbf{u_0} = u_0(y^\prime) \hat{e}_{\tilde{x}}, \, \rho_0 = \rho_0(y^\prime), \,\mathbf{B_0} = B_0(y^\prime) \hat{e}_{x^\prime} 
\end{equation}
with double-$\tanh$ profiles for $u_0$ and $\rho_0$:
\begin{subequations}
    \begin{equation}
        u_0(y^\prime) = \left\{ \begin{array}{lr}
        \frac{1}{2} \left[ \left( u_f + u_s \right) + \left( u_f - u_s \right) \tanh{\left( \frac{y^\prime - \frac{1}{4}L_{y^\prime}}{a} \right)} \right], & \quad y^\prime < \frac{L_{y^\prime}}{2}  \\
       \frac{1}{2} \left[ \left( u_f + u_s \right) - \left( u_f - u_s \right) \tanh{\left( \frac{y^\prime - \frac{3}{4}L_{y^\prime}}{a} \right)} \right], & \quad y^\prime \ge \frac{L_{y^\prime}}{2}
        \end{array}\right.
    \end{equation}
    \begin{equation}
        \rho_0(y^\prime) = \left\{ \begin{array}{lr}
        \frac{1}{2} \left[ \left( \rho_f + \rho_s \right) + \left( \rho_f - \rho_s \right) \tanh{\left( \frac{y^\prime - \frac{1}{4}L_{y^\prime}}{a} \right)} \right], & \quad y^\prime < \frac{L_{y^\prime}}{2}  \\
       \frac{1}{2} \left[ \left( \rho_f + \rho_s \right) - \left( \rho_f - \rho_s \right) \tanh{\left( \frac{y^\prime - \frac{3}{4}L_{y^\prime}}{a} \right)} \right], & \quad y^\prime \ge \frac{L_{y^\prime}}{2}
        \end{array}\right.
    \end{equation}
\end{subequations}
and a uniform magnetic field
\begin{equation}
    B_0(y^\prime) = B_0
\end{equation}
in all the simulations. The width of the shear region is $a=0.075 L_{y^\prime}$ with $L_{y^\prime}$ to be the size of the simulation domain along $\hat{e}_{y^\prime}$. $u_s, u_f, \rho_s, \rho_f$ are the speeds and densities for the slow and fast streams respectively. For all the runs, the initial location of the simulation domain is 
\begin{equation}
    R_0 = 30 R_s = 0.14 \mathrm{AU}
\end{equation}
where $R_s$ is the solar radius and the size of the domain is
\begin{equation}
    L_{x^\prime} \times L_{y^\prime} = 10R_s \times  \pi R_0
\end{equation}
i.e. the domain is a half-circle in the ecliptic plane. The initial spiral angle $\alpha$, if not zero, is set to be 
\begin{equation}
    \alpha = 0.142
\end{equation}
so that at 1 AU the spiral angle is around $\pi/4$, in accordance with the observation. The strength of the magnetic field is $B_0 = 250 \, \mathrm{nT}$ so that at 1 AU $B_r \approx B_\varphi \approx 5 \, \mathrm{nT}$. The densities of the slow and fast streams are $n_s = 360 \, \mathrm{cm}^{-3}$ and  $n_f = 140 \, \mathrm{cm}^{-3}$. The speeds of the slow and fast streams are $u_s = 340 \, \mathrm{km/s}$ and $u_f = 700 \, \mathrm{km/s}$ and the mean radial speed is $U_0 = 464 \, \mathrm{km/s}$. The thermal pressure is $p_0 = 5 \, \mathrm{nPa}$ so that the temperatures of the slow and fast streams are $T_s = 1.0\times 10^6 \, \mathrm{K}$ and $T_f = 2.6\times 10^6 \, \mathrm{K}$. The adiabatic index is $\kappa=3/2$ instead of $\kappa = 5/3$ to prevent the plasma from cooling down too fast. Note that the radial decay of the temperature due to expansion obeys $T \propto R^{-2(\kappa-1)}$ so that with $\kappa = 3/2$ the temperatures of the slow and fast streams at 1 AU are $T_s = 1.4 \times 10^5 \, \mathrm{K}$ and $T_f = 3.6 \times 10^5 \, \mathrm{K}$. The normalization units are: $\Bar{B} = 250 \, \mathrm{nT}$, $\Bar{n} = 200 \mathrm{cm}^{-3}$ and $\Bar{L}= R_s$ which lead to the unit speed $\Bar{U} = \Bar{B}/\sqrt{\mu_0 m_i \Bar{n}} = 385.6 \, \mathrm{km/s}$ and unit pressure $\Bar{p}= \Bar{n}m_i \Bar{U}^2 = 49.7 \, \mathrm{nPa}$ where $m_i$ is the proton mass.

We add circularly-polarized Alfv\'enic wave bands on top of the stream structure:
\begin{subequations}
    \begin{equation}
        \mathbf{b_{1,o}} = \delta b \sum_{N=1}^{N_{max}} \frac{1}{\sqrt{N}} \left[\cos\left(\frac{2\pi N}{L_{x^\prime} }x^\prime + \phi_{N,o} \right) \hat{e}_{y^\prime} +  \sin \left(\frac{2\pi N}{L_{x^\prime} }x^\prime + \phi_{N,o} \right) \hat{e}_{z} \right], \quad \mathbf{u_{1,o}} = -\frac{\mathbf{b_{1,o}}}{\sqrt{\rho_0(y^\prime)}}
    \end{equation}
    \begin{equation}
        \mathbf{b_{1,i}} = r_{io}\times \delta b \sum_{N=1}^{N_{max}} \frac{1}{\sqrt{N}}\left[\cos\left(\frac{2\pi N}{L_{x^\prime} }x^\prime + \phi_{N,i} \right) \hat{e}_{y^\prime} +  \sin \left(\frac{2\pi N}{L_{x^\prime} }x^\prime + \phi_{N,i} \right) \hat{e}_{z} \right], \quad \mathbf{u_{1,i}} = \frac{\mathbf{b_{1,i}}}{\sqrt{\rho_0(y^\prime)}}
    \end{equation}
\end{subequations}
Here $\delta b$ is the amplitude of the magnetic perturbation of the outward wave, $r_{io}$ is the ratio between the amplitudes of inward and outward waves, $\phi_{N,o}$ and $\phi_{N,i}$ are the random phases of mode $N$ of outward and inward waves. The slope of the power spectrum of the wave band is $-1$. In order to make sure $\nabla \cdot \mathbf{b_{1}} = 0$, $\mathbf{b_{1}}$ is invariant along $y^\prime$ and $\mathbf{u_{1}}$ varies with $y^\prime$ due to the non-uniform density. This leads to the inhomogeneity of the Alfv\'en wave energy along $y^\prime$ direction: the wave energy is larger in the fast stream than the slow stream. Five 2D runs are carried out and they are listed in Table \ref{tab:simulation_parameters}. By choosing the parameter $\delta b$, the total energies in the waves are invariant among the runs. We fix $N_{max} = 16 $ in all the simulations. The maximum simulation time is $t=200$, corresponding to a radial distance $R=270.9 R_s =1.26 \, \mathrm{AU}$. The resolution is $n_{x^\prime}\times n_{y^\prime} = 2048 \times 4096$. In addition, we also make a 1D run ($L_{y^\prime} = \pi R_0$ and $n_{y^\prime} = 1024$) without adding waves to show the evolution of the stream structure up to $R=400 R_s$.

\begin{table}
\centering
\begin{tabular}{l|llll}
\hline
Run & Expansion & Corotation &$r_{io}$  & $\delta b$  \\
\hline
A0 & N  & N & 0.2 & 0.2 \\
A & Y  & N & 0.2 & 0.2 \\
B & Y  & Y & 0.2 & 0.2 \\
C & Y  & Y & 1.0 & 0.144 \\
D & Y  & Y & 5.0 & 0.04 \\
\hline
\end{tabular}
\caption{Parameters of the 2D runs. Here $r_{io}$ is the ratio between the amplitude of the inward waveband and the amplitude of the outward waveband. $\delta b$ is the amplitude of the outward waveband and the five runs have the same total wave energies. If expansion is present, the radial mean speed $U_0=1.2$. If corotation is present, the initial spiral angle $\alpha = 0.142$. }\label{tab:simulation_parameters}
\end{table}

\section{Results}\label{sec:results}
\subsection{1D Run without Waves}

\begin{figure}[ht!]
\centering
\includegraphics[scale=0.65]{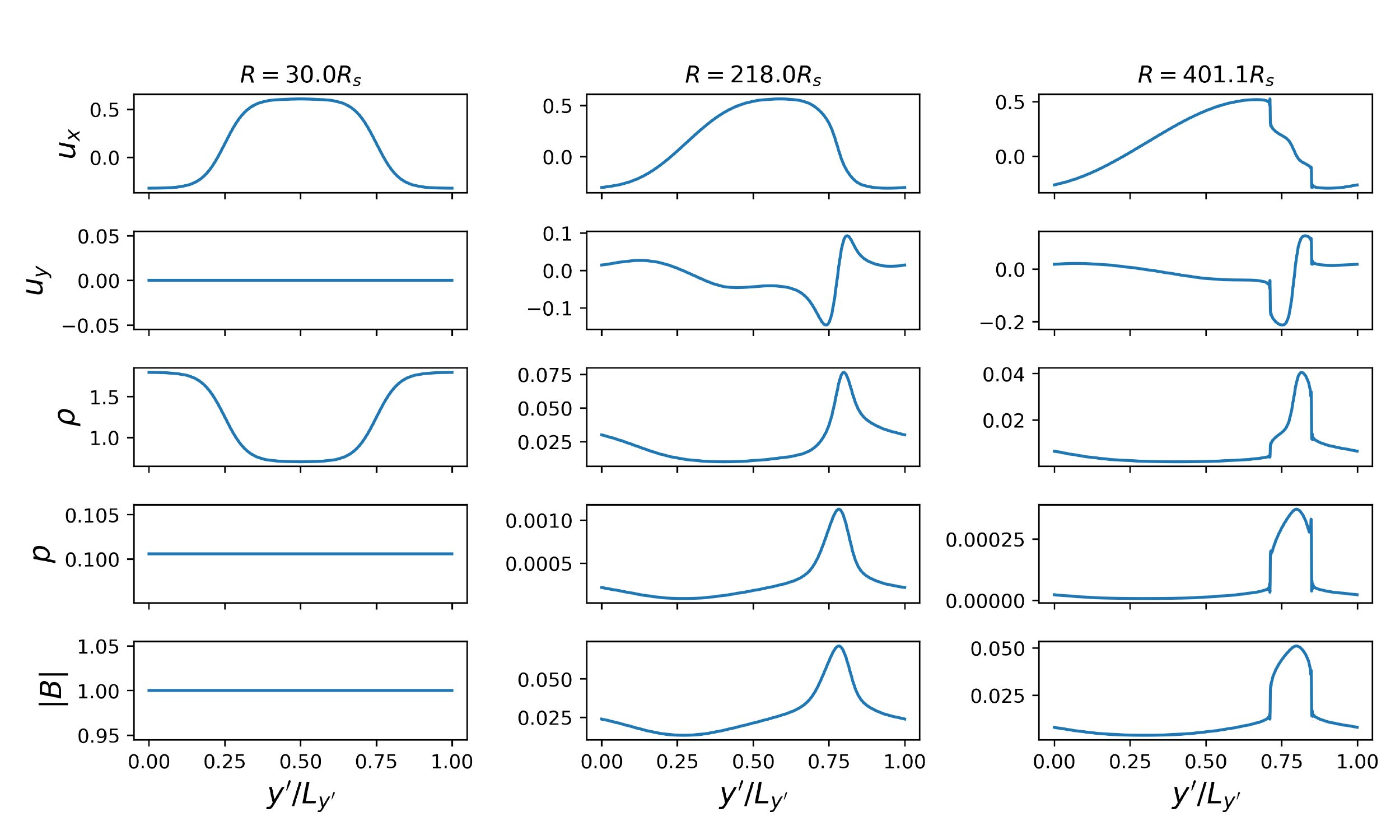}
\caption{1D simulation of the large-scale stream structure. From top to bottom rows are longitudinal ($y^\prime$) profiles of the radial velocity, the azimuthal velocity, the density, the pressure and the magnitude of magnetic field respectively. From left to right columns are the snapshots at $R=30.0R_s, \, 218.0 R_s, \, 401.1 R_s$. \label{fig:1D_sim}}
\end{figure}

In this section we show a 1D test simulation of the stream structure without adding any waves. This run serves as a test of the code. For convenience, we refer to $y^\prime / L_{y^\prime}$ as the normalized ``longitude'' hereinafter although $\hat{e}_{y^\prime}$ is not exactly along the azimuthal direction $\hat{e}_y$. Figure \ref{fig:1D_sim} shows the radial evolution of the longitudinal profiles of the radial velocity $u_x$, the azimuthal velocity $u_y$, the density $\rho$, the pressure $p$ and the magnitude of the magnetic field $|B|$ (from top to bottom rows). The left, middle and right columns are snapshots at $R=30.0R_s, \, 218.0R_s \, \& \, 401.1R_s$ respectively. At around 1 AU (middle column), a clear compression region already forms. The flows are deflected away from the interface between fast and slow streams. The density, pressure and magnetic field peak around the compression region. Further out, a forward-backward shock pair, which bounds the compression region, forms as shown in the right column. The results, shown in Figure \ref{fig:1D_sim}, are consistent with \citep{grappin1996} and may be benchmarked against their Figure 3.

\subsection{Diagnostics of the Alfv\'enic turbulence}
Before presenting the results of the 2D simulations, we first introduce the diagnostics adopted for the analysis of the simulation data.

The analysis is mainly based on the perturbed Els\"asser variables $\mathbf{z_{out}}$ and $\mathbf{z_{in}}$. The procedure to calculate them is described as follows. We first calculate the $x^\prime$-averaged, i.e. the background, magnetic and velocity fields: 
\begin{equation}
    \mathbf{B_0}(y^\prime) = \frac{1}{L_{x^\prime}} \int_{0}^{L_{x^\prime}} \mathbf{B}(x^\prime,y^\prime) dx^\prime , \quad \mathbf{u_0}(y^\prime) = \frac{1}{L_{x^\prime}} \int_{0}^{L_{x^\prime}} \mathbf{u}(x^\prime,y^\prime) dx^\prime 
\end{equation}
and then the perturbed magnetic and velocity fields:
\begin{equation}
    \mathbf{b_1}(x^\prime, y^\prime) = \mathbf{B}(x^\prime, y^\prime) - \mathbf{B_0}(y^\prime), \quad  \mathbf{u_1}(x^\prime, y^\prime) = \mathbf{u}(x^\prime, y^\prime) - \mathbf{u_0}(y^\prime)
\end{equation}
The Els\"asser variables are then calculated by 
\begin{equation}\label{eq:definition_zin_zout}
    \mathbf{z_{out}} = \mathbf{u_1} - \mathrm{sign}(B_{0x}) \frac{\mathbf{b_1}}{\sqrt{\rho}}, \quad \mathbf{z_{in}} = \mathbf{u_1} + \mathrm{sign}(B_{0x}) \frac{\mathbf{b_1}}{\sqrt{\rho}}
\end{equation}
where $\mathrm{sign}(B_{0x})$ is the sign of the radial background magnetic field. Note that the density is not $x^\prime$-averaged but the local density. We further project the Els\"asser variables defined by Eq (\ref{eq:definition_zin_zout}) into three directions: the out-of-plane direction $\hat{e}_z$, the parallel-to-$\mathbf{B_0}$ direction $\hat{e}_{x^\prime}$, and the in-plane perpendicular-to-$\mathbf{B_0}$ direction $\hat{e}_{\perp} = \hat{e}_z \times \hat{e}_{x^\prime}$. In the analysis hereinafter, we only deal with the $z$-component and the perpendicular component and exclude the parallel component. At a certain time $t$, various energies as functions of $y^\prime$ are calculated by integrating along the $x^\prime$ direction, e.g. the outward Els\"asser energy:
\begin{equation}
    E_{out}(y^\prime,t) = \frac{1}{2} \int_{x^\prime} \left( z_{out,z}^2 + z_{out,\perp}^2 \right)
\end{equation}
The total energy, the normalized cross helicity and the normalized residual energy are then calculated by 
\begin{equation}
    E^T = E_{out} + E_{in}, \quad \sigma_c = \frac{E_{out} - E_{in}}{E_{out} + E_{in}}, \quad \sigma_r = \frac{E_u - E_b}{E_u + E_b}
\end{equation}
The kinetic and magnetic energies are those in the perturbations $\mathbf{u_1}$ and $\mathbf{b_1}$ and we do not include the parallel component in calculating $\sigma_r$. We have verified that including the parallel component in $E_u$ and $E_b$ does not make a significant difference. The normalized density perturbation $\delta \rho/ \rho$ is the root-mean-square value of $\rho$ along $x^\prime$ divided by the $x^\prime$-averaged density $\rho_0$:
\begin{equation}
    \frac{\delta \rho}{\rho} (y^\prime,t) = \frac{1}{\rho_0(y^\prime,t)}  \sqrt{ \frac{1}{L_{x^\prime}} \int_{x^\prime} \left[\rho(x^\prime,y^\prime,t) - \rho_0(y^\prime,t) \right]^2}
\end{equation}
Power spectra of $\mathbf{z_{out}}$ and $\mathbf{z_{in}}$ are calculated along $\hat{e}_{x^\prime}$ and $\hat{e}_{y^\prime}$ by applying Fourier transform to the $z$ and perpendicular components of them, e.g. $E_{out,z} \left( k_{x^\prime}, y^\prime, t \right) =  \left| \mathcal{F}_{x^\prime} \left(z_{out,z} \right)\right|^2$ where $\mathcal{F}_{x^\prime}$ is the Fourier transform in coordinate $x^\prime$. When we present the spectra, we further average the spectra along the non-Fourier-transformed coordinates to eliminate the strong oscillations. The details of the averaging procedure of the spectra will be discussed later.

\subsection{Run A0: no corotation, no expansion, outward-dominant waves}

\begin{figure}[ht!]
\centering
\includegraphics[scale=0.6]{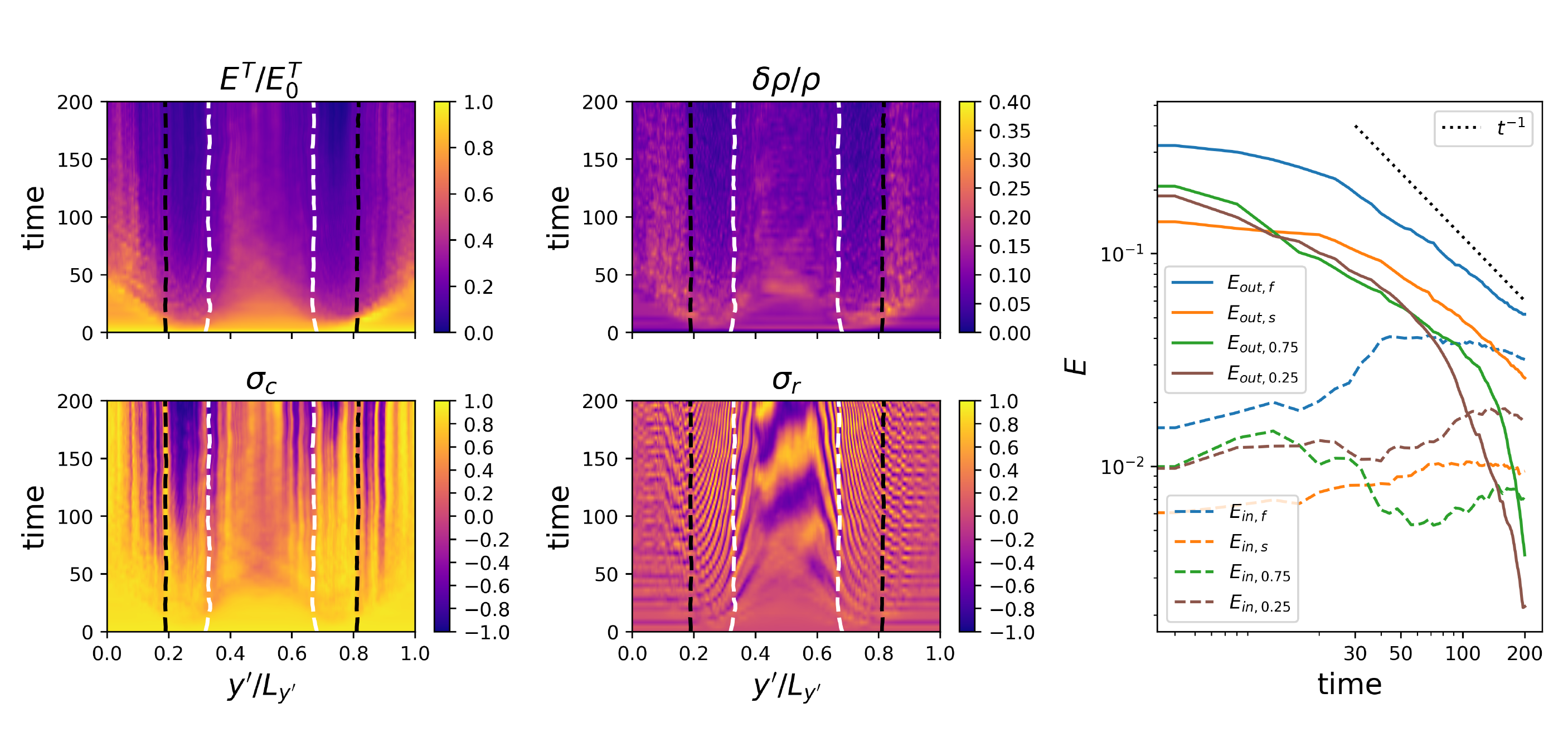}
\caption{Results of Run A0. The left two columns are the $y^\prime-t$ contours of the total Els\"asser energy $E^T/E_0^T$ where $E_0^T$ is $E^T(t=0)$ (top-left), the normalized density perturbation $\delta \rho /\rho$ (top-middle), the normalized cross-helicity $\sigma_c$ (bottom left) and the normalized residual energy $\sigma_r$ (bottom middle). White dashed lines mark the longitudes where the $x^\prime$-averaged radial speed $u_{0x}$ equals $650 \, \mathrm{km/s}$ and black dashed lines mark the longitudes where the $x^\prime$-averaged radial speed  $u_{0x}$ equals $400 \, \mathrm{km/s}$. Right panel shows the time evolution of the Els\"asser energies in log-log scale where solid/dashed curves are the outward/inward waves and blue, orange, green and brown represent fast stream (``f''), slow stream (``s''), shear region around $y^\prime=0.75L_{y^\prime}$ and shear region around $y^\prime=0.25L_{y^\prime}$. The black dotted line is $E \propto t^{-1}$ for reference. \label{fig:ET_rms_rho_sigma_c_sigma_r_Phi_Rt_RunA0}}
\end{figure}

In Run A0, the background fields are radial, i.e. there is no compression and rarefaction. Besides, the expansion effect is turned off. The initial condition consists of the outward-dominant Alfv\'en wave band. The result of Run A0 is shown in Figure \ref{fig:ET_rms_rho_sigma_c_sigma_r_Phi_Rt_RunA0}. 

The top-left panel shows the $y^\prime-t$ contour of the total Els\"asser energy $E^T/E_0^T$ where $E_0^T$ is $E^T(t=0)$. The white dashed lines mark $y^\prime$ where $u_{0x}$ equals $650$ km/s and the black dashed lines mark $y^\prime$ where $u_{0x}$ equals $400$ km/s (the same in the other three contours). We see that the total Els\"asser energy $E^T$ at all longitudes decays with time while in the shear region the energy decays much faster. The right panel of Figure \ref{fig:ET_rms_rho_sigma_c_sigma_r_Phi_Rt_RunA0} shows the time evolution of the Els\"asser energies of the outward wave (solid curves) and inward wave (dashed curves) averaged in different regions bounded by the white and black dashed lines in the contours, i.e. the fast stream (blue), the slow stream (orange), the shear region around $y^\prime=0.75L_{y^\prime}$ (green) and the shear region around $y^\prime=0.25L_{y^\prime}$ (brown). The black dotted line is $E \propto t^{-1}$ for reference. The evolution of the wave energies inside the fast and slow streams is similar: the outward wave energy decays with time at a rate slower than $t^{-1}$ and the inward wave energy increases with time slightly. Inside the shear regions, the outward wave energy decays slower than $t^{-1}$ first and the decay rate is similar to that of the outward wave inside the fast/slow streams. However, after some time ($t \gtrsim 100 $ in the shear region around $y^\prime=0.75L_{y^\prime}$ and $t \gtrsim 70$ in the shear region around $y^\prime = 0.25L_{y^\prime}$) the wave energy starts to drop very fast. The inward wave energy grows slowly at the beginning, followed by a drop at $t\approx 30$ and then starts to grow again in the two shear regions. Note that in the shear region at $y^\prime \approx 0.75L_{y^\prime}$, the drop of the inward wave energy is stronger than that in the shear region at $y^\prime \approx 0.25L_{y^\prime}$. Here we must point out that the initial configuration, although symmetric in $y^\prime$, does not evolve symmetrically because the $y^\prime$-gradients of the background fields are of opposite signs while the initial perturbations along $y^\prime$ (e.g. $u_{1y^\prime}$) do not change sign at the two shear regions. This, for example, will lead to an increase of $\rho$ at one shear region and a decrease of $\rho$ at the other one.

The top-middle panel of Figure \ref{fig:ET_rms_rho_sigma_c_sigma_r_Phi_Rt_RunA0} shows the $y^\prime-t$ contour of the relative density fluctuation $\delta \rho / \rho$. The value of $\delta \rho / \rho$ remains small ($\lesssim 0.2$) throughout the simulation. The largest density fluctuation $\delta \rho$ is found to be inside the slow stream near the boundaries of the shear regions, as can be seen from the contour. The bottom-left panel displays the $y^\prime-t$ contour of the normalized cross-helicity $\sigma_c$, which decays with distance in all the flow regions. The decay rate is largest inside the shear region at $y^\prime \approx 0.25L_{y^\prime}$ where $\sigma_c$ almost reaches $-1$ at the end of the simulation. This can also be seen from the right column of Figure \ref{fig:ET_rms_rho_sigma_c_sigma_r_Phi_Rt_RunA0} which shows that the outward Els\"asser energy is one order of magnitude smaller than the inward energy in the shear region at $y^\prime \approx 0.25L_{y^\prime}$ at the end of the simulation. A notable phenomenon is the stripe structures in the contour of $\sigma_c$, showing that $\sigma_c$ decays much faster within some narrow channels in $y^\prime$ compared to the ambient streams. The evolution of $\sigma_c$ we find is very similar to that in the incompressible simulation by \citet{roberts1992} (see their Figure 12). The bottom-middle panel shows the $y^\prime-t$ contour of the normalized residual energy $\sigma_r$. Strong oscillations are observed. On average $\sigma_r$ is 0 but the instant amplitude can be as large as $\simeq 1$. The oscillation of $\sigma_r$ is strongest inside the shear regions due to the large longitudinal gradient of the relative speed between the counter-propagating waves. A trend of increasing of $\left| \sigma_r \right|$ in some regions, e.g. the fast stream and the shear regions, is also seen. This is because of the decrease of $\left| \sigma_c \right|$: if $\sigma_c = 0 $, the inward and outward waves are of the same amplitude and thus $\left| \sigma_r \right|$ will be equal to 1 if the two populations of waves are non-correlated. Actually, we can see that at $y^\prime \approx 0.25 L_{y^\prime}$, $\left| \sigma_r \right|$ increases at $t \lesssim 150$ and then starts to drop, which is anti-correlated with $\left| \sigma_c \right|$.

\subsection{Run A: no corotation, expansion, outward-dominant waves}

\begin{figure}[ht!]
\centering
\includegraphics[scale=0.6]{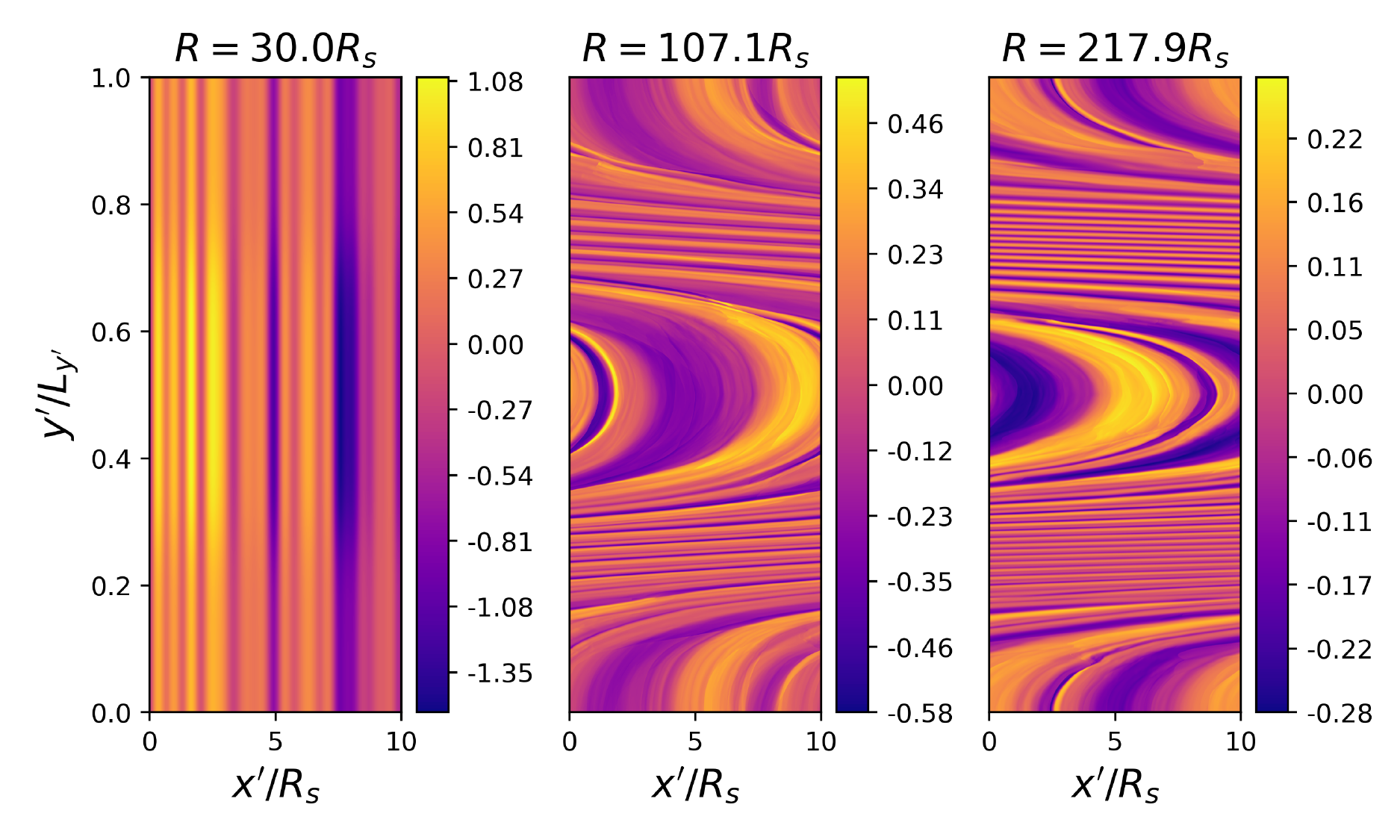}
\caption{Contours of the $z$-component of the outward Els\"asser variable $z_{out,z}$ at $R=30.0R_s, \, 107.1R_s \, \& \,217.9R_s$ in Run A. \label{fig:Run_A_zout_z_contours}}
\end{figure}

\begin{figure}[ht!]
\centering
\includegraphics[scale=0.6]{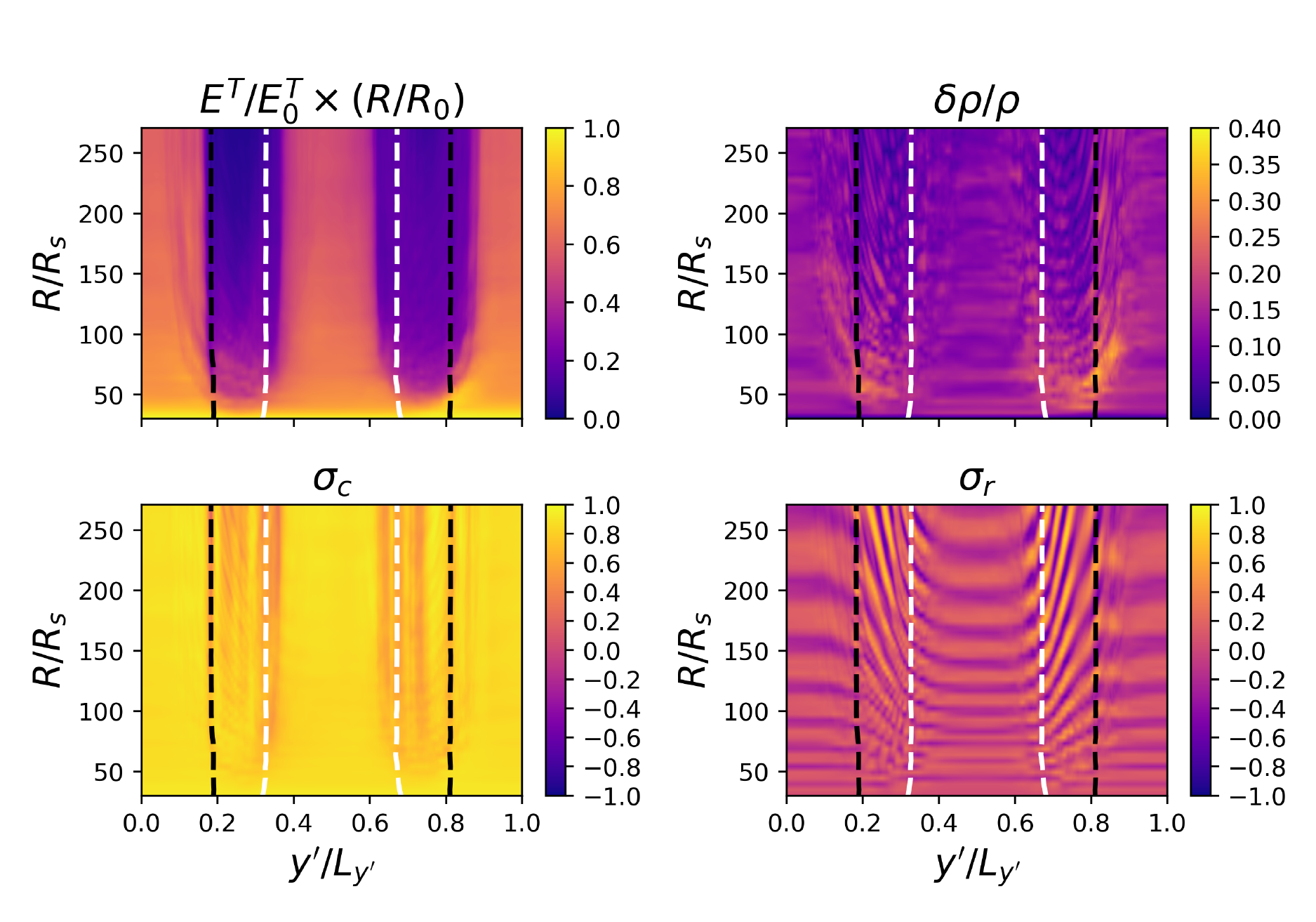}
\caption{$y^\prime-R$ contours of the total Els\"asser energy $E^T/E_0^T$ compensated by $ R/R_0$ where $E_0^T$ is $E^T(t=0)$ (top-left), the normalized density perturbation $\delta \rho /\rho$ (top-right), the normalized cross-helicity $\sigma_c$ (bottom left) and the normalized residual energy $\sigma_r$ (bottom right) for Run A. White dashed lines mark the longitudes where the $x^\prime$-averaged radial speed $u_{0x}$ equals $650 \, \mathrm{km/s}$ and black dashed lines mark the longitudes where the $x^\prime$-averaged radial speed  $u_{0x}$ equals $400 \, \mathrm{km/s}$. \label{fig:ET_rms_rho_sigma_c_sigma_r_Phi_Rt_RunA}}
\end{figure}

In this subsection, we present the results of Run A ($\alpha = 0$, $r_{io}=0.2$ and $\delta b = 0.2$), where the compression between the fast and slow streams is absent but the expansion effect is turned on.  

Figure \ref{fig:Run_A_zout_z_contours} shows the contours of the out-of-plane component of the outward Els\"asser variable $z_{out,z}$ at three radial distances: $R=30.0R_s, \, 107.1R_s \, \& \, 217.9R_s$. From Figure \ref{fig:Run_A_zout_z_contours}, it is clearly seen that the differential radial flow leads to the phase-mixing of the Alfv\'en waves, the wave vector of which is tilted from $\hat{e}_{x^\prime}$ toward $\hat{e}_{y\prime}$. The strongest phase-mixing happens in the regions where the velocity shear is the largest (around $y^\prime =0.75 L_{y^\prime} $ and $y^\prime = 0.25 L_{y^\prime}$). The dissipation of waves is observed at these regions since the phase-mixing transfers the wave energy to small scales where the numerical dissipation is strong. For other Els\"asser variables, i.e. $z_{out,\perp}$, $z_{in,z}$ and $z_{in,\perp}$, similar evolution is also observed.

Figure \ref{fig:ET_rms_rho_sigma_c_sigma_r_Phi_Rt_RunA} displays the $y^\prime-R$ contours of the total Els\"asser energy $E^T/E^T_0$ compensated by $R/R_0$ where $E^T_0$ is $E^T$ at $t=0$ (top-left), the relative density fluctuation $\delta \rho / \rho$ (top-right), the normalized cross-helicity $\sigma_c$ (bottom-left) and the normalized residual energy $\sigma_r$ (bottom-right), similar to Figure \ref{fig:ET_rms_rho_sigma_c_sigma_r_Phi_Rt_RunA0} but note that the y-axis is now radial distance instead of time. The white dashed lines mark $y^\prime(R)$ where the $x^\prime$-averaged radial speed $u_{0x}$ equals $650 \, \mathrm{km/s}$ and the black dashed lines mark $y^\prime(R)$ where $u_{0x}$ equals $400 \, \mathrm{km/s}$. The decay of $E^T$ is in general faster than $1/R$, the WKB prediction of the Alfv\'en waves in the spherical geometry \citep{belcher1971}. Similar to Run A0, it clearly shows a longitudinal variation: inside the fast and slow streams, the decay is slower than in the shear regions. The relative density fluctuation $\delta \rho / \rho$ is smaller than $0.2$ most of the time and it is smaller inside the shear regions compared with the fast and slow streams. It is also observed that some density structures are generated near the boundaries between the shear regions and the slow stream and propagate along the $y^\prime$ direction. The most significant one starts at $R\approx 60 R_s$ and $y^\prime \approx 0.8 L_{y^\prime}$, with amplitude $\delta \rho / \rho \approx 0.35$. Note that in Run A0 we also observe that the density fluctuation is largest near the boundary between the shear region and the slow stream. 

It is known from the observations that the normalized cross-helicity decreases with radial distance \citep[e.g.][]{roberts1987a,roberts1987b}. The possible mechanisms for the decrease include the generation of inward Alfv\'en waves due to the velocity shears and the faster decay of outward Alfv\'en waves with distance compared with the inward waves \citep{bruno1991}. In Run A0 we already see that the velocity shear leads to the drop of $\sigma_c$. From Figure \ref{fig:ET_rms_rho_sigma_c_sigma_r_Phi_Rt_RunA}, we confirm that $\sigma_c$ drops with radial distance inside the shear regions, especially near the boundaries of the fast stream. It decreases to values around $0.7-0.8$ within $100 R_s$ and then decreases slowly to around $ 0.5-0.6$ until the end of the simulation $R=270.9R_s$. In the fast and slow streams, $\sigma_c$ remains almost constant around the initial value $0.92$. Compared with Run A0, the contour of $\sigma_c$ is quite smooth and no stripe-like structures are formed, indicating that the expansion effect slows down the evolution of the wave energies. Last, we look at the residual energy shown in the bottom-right panel. Similar to Run A0, the normalized residual energy fluctuates around $0$ and no systematic growth of $\sigma_r$ is observed. However, the oscillation of $\sigma_r$ is much weaker in Run A than in Run A0 because the expansion reduces the Alfv\'en speed so that the relative speed between the outward and inward waves goes down with radial distance.

\begin{figure}[ht!]
\centering
\includegraphics[scale=0.5]{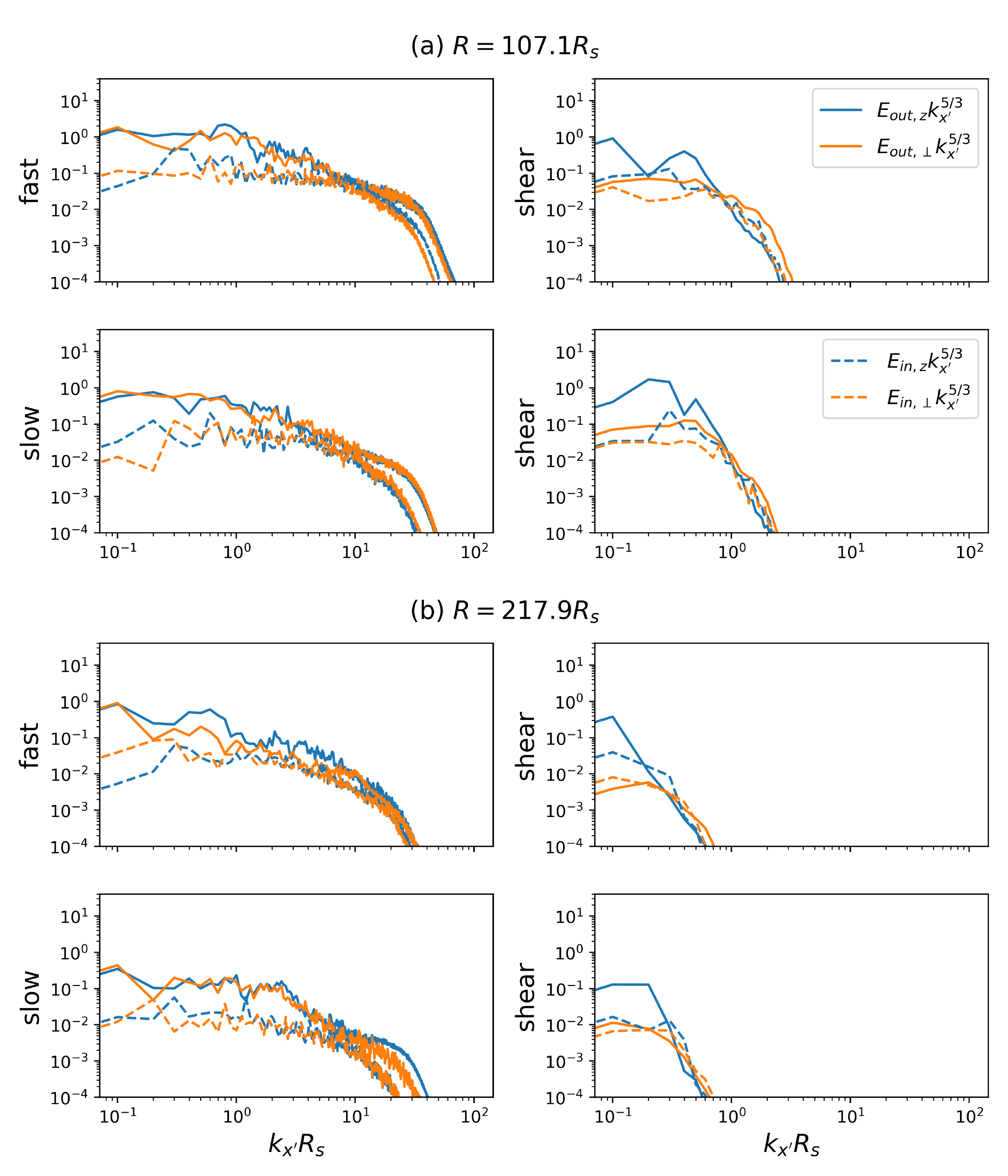}
\caption{Power spectra of the Els\"asser variables in Run A at (a) $R=107.1R_s$ and (b) $R=217.9R_s$ averaged in different regions: fast stream (top left), slow stream (bottom left), shear region at $y^\prime=0.75 L_{y^\prime}$ (top right) and shear region at $y^\prime=0.25 L_{y^\prime}$ (bottom right). The spectra are calculated along parallel direction $\hat{e}_{x^\prime}$. Blue and orange solid lines are the $z$-component and in-plane perpendicular-to-$\mathbf{B_0}$-component of the outward Alfv\'en wave. Blue and orange dashed lines are the two components of the inward Alfv\'en wave. The spectra are multiplied by $k_{x^\prime}^{5/3}$.\label{fig:spectrum_zinzout_by_regions_RunA}}
\end{figure}

Figure \ref{fig:spectrum_zinzout_by_regions_RunA} shows the power spectra of the Els\"asser variables along the parallel direction $\hat{e}_{x^\prime}$ (in this run it is aligned with the radial direction) at (a) $R=107.1 R_s\approx 0.5 \, \mathrm{AU}$ and (b) $R=217.9R_s \approx 1 \, \mathrm{AU}$. Again we divide the domain into four regions: the fast stream, the slow stream and the two shear regions (in Run B-D they are the compression/rarefaction regions). The spectra in different regions are displayed in four subplots at each time. The shear region plotted on the top row is the one at $y^\prime \approx 0.75 L_{y^\prime}$. The spectra are averaged in $y^\prime$ inside each region and are multiplied by $k_{x^\prime}^{5/3}$. The blue and orange solid curves are the $z$ and $\perp$ components of the outward Alfv\'en waves and the dashed curves are of the inward waves. Inside the shear regions, the wave energies are strongly damped and inertial ranges are not observed in the spectra and as the radial distance increases, the spectra are eroded rapidly. In the fast and slow streams, the spectra behave similarly and are more stable compared with the shear regions. Especially, $\mathbf{z_{out}}$ shows clear 3-segment spectra: the large scales with $k_{x^\prime}R_s \lesssim 1$, the intermediate scales with $1 \lesssim k_{x^\prime}R_s \lesssim 20-30$ and the small scales which are dominated by the numerical dissipation. At 0.5 AU, the large scales show slopes close to $-5/3$ in both the fast and slow streams while at 1 AU the large-scale part of the $\mathbf{z_{out}}$ spectrum in the fast stream is eroded by the intermediate-scale part and steepening of the spectrum is observed.

\begin{figure}[ht!]
\centering
\includegraphics[scale=0.6]{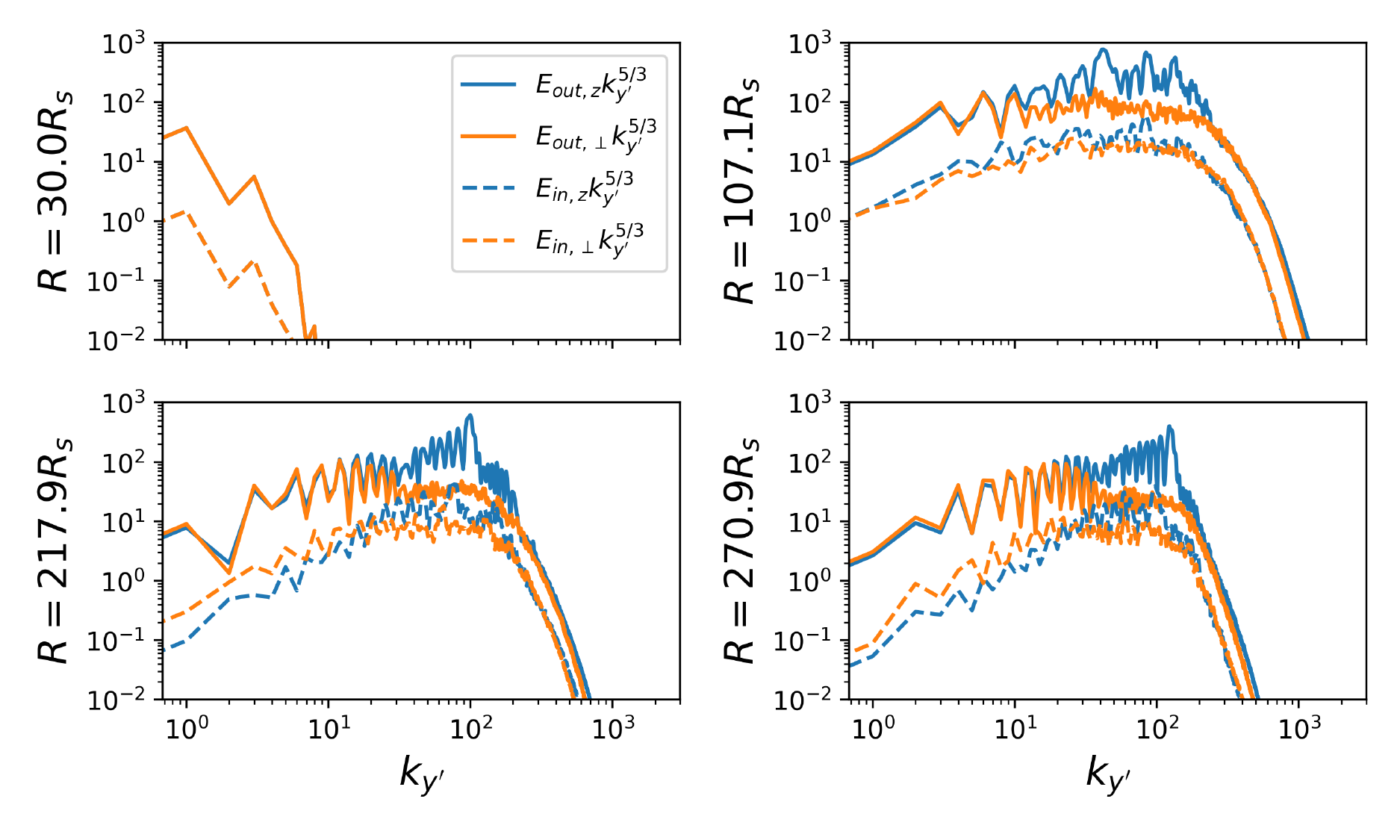}
\caption{Power spectra of the Els\"asser variables calculated along $\hat{e}_{y^\prime}$ and averaged in $x^\prime$ in Run A. From top-left to bottom-right are at $R=30.0R_s, \, 107.1R_s, \, 217.9R_s \, \& \, 270.9R_s$ respectively. Blue and orange solid lines are the $z$-component and in-plane perpendicular-to-$\mathbf{B_0}$-component of the outward Alfv\'en wave. Blue and orange dashed lines are the two components of the inward Alfv\'en wave. The spectra are multiplied by $k_{y^\prime}^{5/3}$. \label{fig:spectrum_zinzout_along_y_RunA}}
\end{figure}

Figure \ref{fig:spectrum_zinzout_along_y_RunA} shows the power spectra of the Els\"asser variables along $\hat{e}_{y^\prime}$ direction at four radial distances: $R=30.0R_s, \, 107.1R_s, \, 217.9R_s, \, 270.9R_s$. The wavenumber $k_{y^\prime}$ is defined by the normalized $y^\prime$, i.e. $y^\prime/L_{y^\prime}$, so that $k_{y^\prime} \in [0,n_{y^\prime}/2]$. The blue and orange lines are the $z$ and $\perp$ components of the outward Alfv\'en wave and the dashed lines are those of the inward Alfv\'en wave. The spectra are averaged in $x^\prime$ and multiplied by $k_{y^\prime}^{5/3}$. At $R=30.0R_s$, i.e. the initial state, the curves for $z$-components are covered by those of $\perp$-components as the initial wave band is circularly-polarized. Kolmogorov-like inertial range which spans about one decade forms at 0.5 AU for all the wave components. It maintains throughout the simulation for $E_{out,\perp}$ and $E_{in,\perp}$. But for $E_{in,z}$ the inertial range shortens with radial distance and for $E_{out,z}$ the inertial range becomes shallower than $k_{y^\prime}^{-5/3}$ at 1 AU. This asymmetry between the $\perp$-component and the $z$-component is due to the uniformity in $z$-direction which rules out the nonlinear interaction between the waves along $\hat{e}_z$.

\subsection{Run B: corotation, expansion, outward-dominant waves}
In this section we present the results of Run B ($\alpha=0.142$, $r_{io}=0.2$ and $\delta b = 0.2$). This run has the most realistic setup: expansion, velocity shear and compression/rarefaction are all present and the initial perturbations are outward-dominant Alfv\'en waves.

\begin{figure}[ht!]
\centering
\includegraphics[scale=0.6]{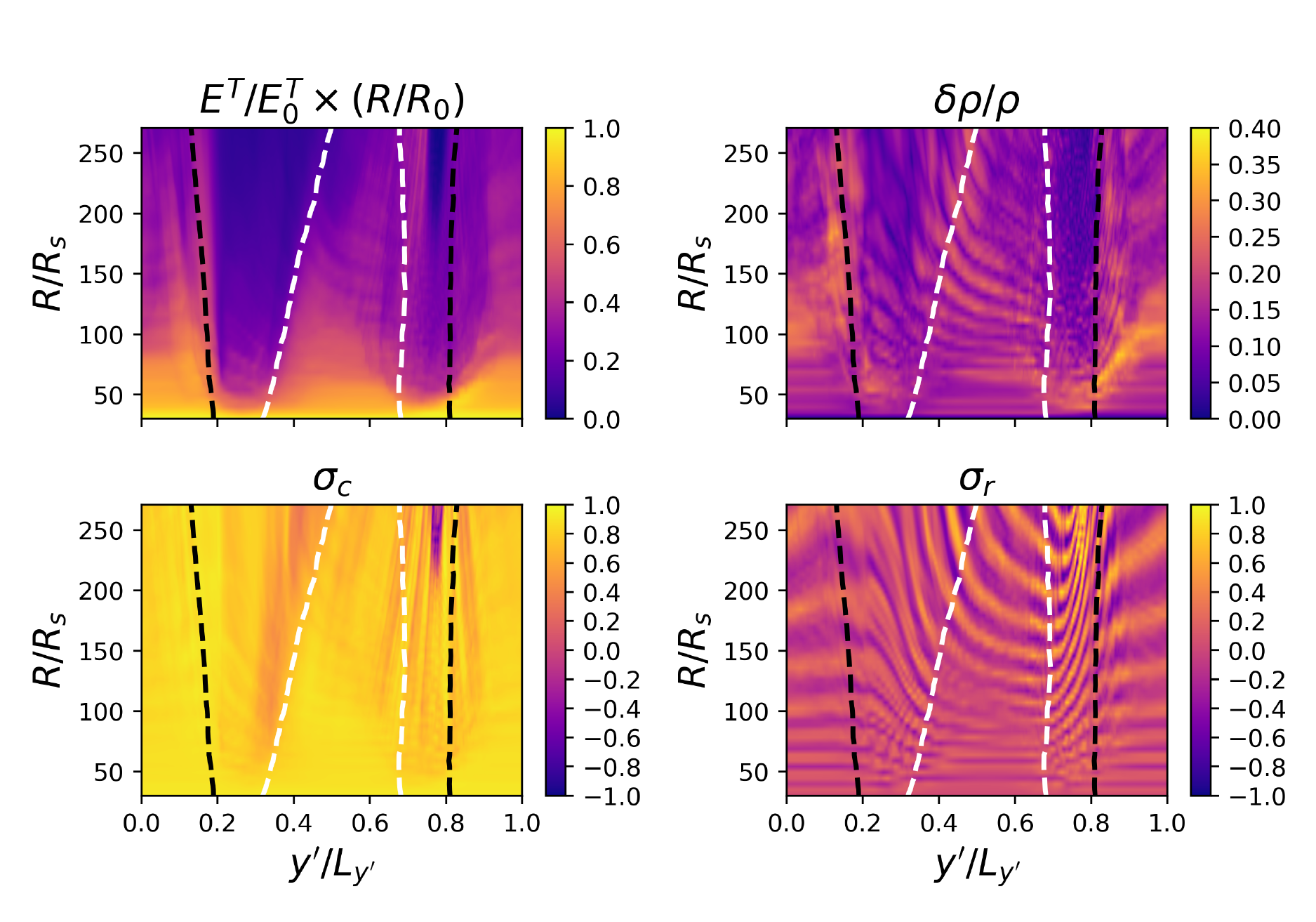}
\caption{$y^\prime -R$ contours of the total Els\"asser energy $E^T/E_0^T$ compensated by $ R/R_0$ where $E_0^T$ is $E^T(t=0)$ (top-left), the normalized density perturbation $\delta \rho /\rho$ (top-right), the normalized cross-helicity $\sigma_c$ (bottom left) and the normalized residual energy $\sigma_r$ (bottom right) for Run B. White dashed lines mark the longitudes where the $x^\prime$-averaged radial speed $u_{0x}$ equals $650 \, \mathrm{km/s}$ and black dashed lines mark the longitudes where the $x^\prime$-averaged radial speed  $u_{0x}$ equals $400 \, \mathrm{km/s}$. \label{fig:ET_rms_rho_sigma_c_sigma_r_Phi_Rt_RunB}}
\end{figure}

Figure \ref{fig:ET_rms_rho_sigma_c_sigma_r_Phi_Rt_RunB} is a similar plot with Figure \ref{fig:ET_rms_rho_sigma_c_sigma_r_Phi_Rt_RunA} for Run B. From top-left to bottom right are the corrected total Els\"asser energy $E^T/E^T_0 \times (R/R_0)$, the normalized density fluctuation, the normalized cross-helicity and the normalized residual energy. The white and black dashed lines mark $u_{0x} = 650 \, \mathrm{km/s}$ and $u_{0x} = 400 \, \mathrm{km/s}$ respectively. Similar to Run A, The total energy decays faster than the WKB prediction $R^{-1}$. However, in the fast and slow streams, the radial decay of $E^T$ is significantly faster in Run B than in Run A. Besides, in Run B, beyond $R\approx 200 R_s$, a narrow band inside the compression region forms at $y^\prime \approx 0.75 L_{y^\prime}$, where the wave energy is much more damped compared with the shear regions in Run A.
This might be due to the fact that the compression between the fast and slow streams steepens the velocity profile, enlarging the velocity shear. $\delta \rho /\rho$ and $\sigma_r$ do not show significant differences between Run A and Run B. Similar to Run A, the decrease of $\sigma_c$ is more significant in the compression and rarefaction regions than inside the fast and slow streams. In the rarefaction region, mainly in the trailing edge of the fast stream, $\sigma_c$ drops to around $0.6$ very soon at $R \approx 80 R_s$ and remains around this value until the end of the simulation. In the compression region, however, $\sigma_c$ remains relatively large ($ > 0.5$) for a long time followed by a fast drop beyond $R \approx 1 \, \mathrm{AU}$ and reaches around $-0.7$ at the end of the simulation $R=270.9R_s$. The drop of $\sigma_c$ coincides with the drop of $E^T$ in the compression region (see the top-left panel). In the fast and slow streams, $\sigma_c$ decreases with distance more slowly, from the initial value $0.92$ to $\sim 0.7-0.8$ at 1 AU. Note that in Run A, $\sigma_c $ remains almost constant around the initial value $0.92$ inside the fast and slow streams, i.e. the velocity shear only reduces the normalized cross-helicity locally in the shear regions. Thus, the compression between the fast and slow streams might play an important role in the radial evolution of $\sigma_c$. It not only speeds up the drop of $\sigma_c$ in the compression region but also speeds up the decrease of $\sigma_c$ inside the fast and slow streams by steepening the velocity profile at all longitudes.

\begin{figure}[ht!]
\centering
\includegraphics[scale=0.5]{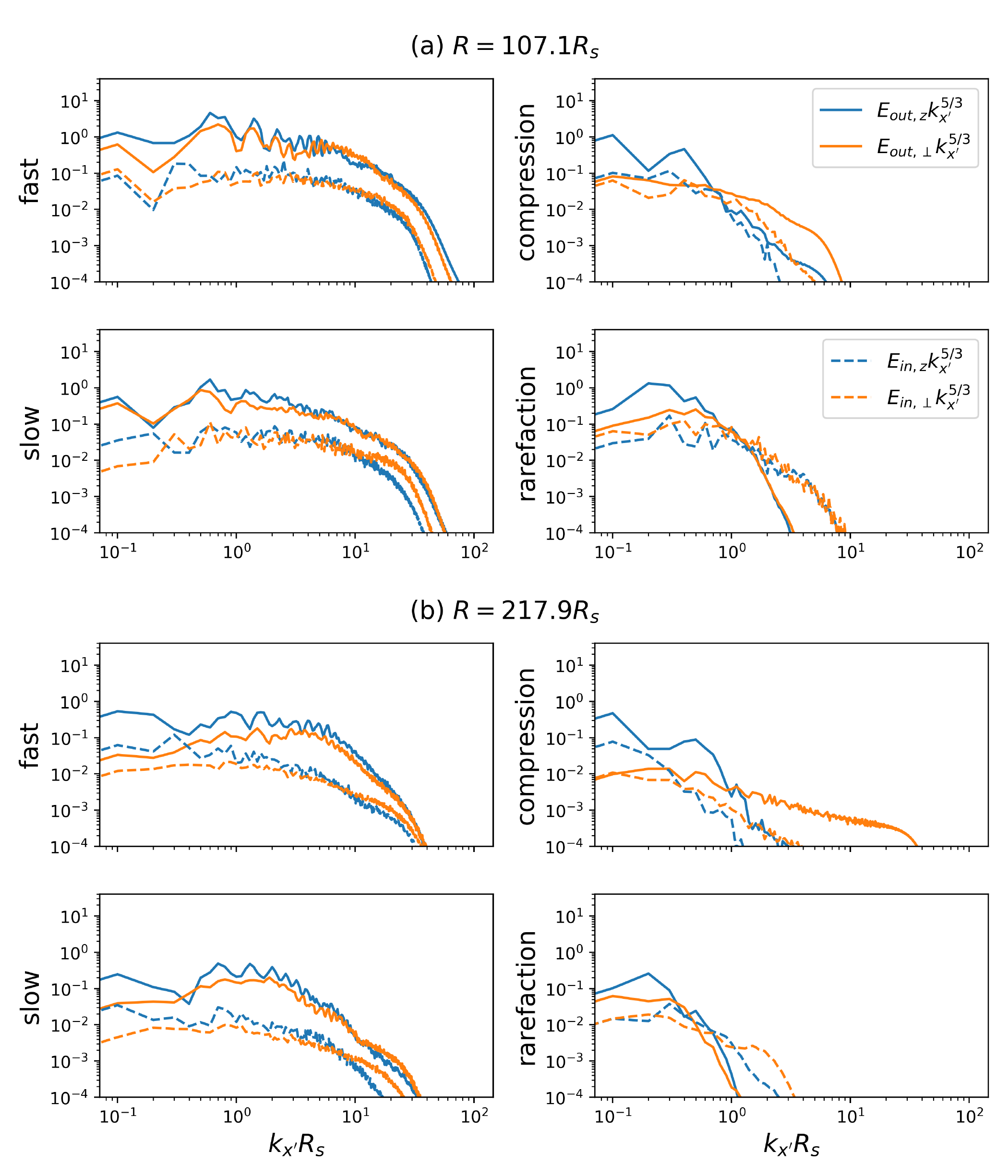}
\caption{Power spectra of the Els\"asser variables in Run B at (a) $R=107.1R_s$ and (b) $R=217.9R_s$ averaged in different regions: fast stream (top left), slow stream (bottom left), compression region (top right) and rarefaction region (bottom right). The spectra are calculated along parallel direction $\hat{e}_{x^\prime}$. Blue and orange solid lines are the $z$-component and in-plane perpendicular-to-$\mathbf{B_0}$-component of the outward Alfv\'en wave. Blue and orange dashed lines are the two components of the inward Alfv\'en wave. The spectra are multiplied by $k_{x^\prime}^{5/3}$. \label{fig:spectrum_zinzout_by_regions_RunB}}
\end{figure}

Figure \ref{fig:spectrum_zinzout_by_regions_RunB} shows the $k_{x^\prime}^{5/3}$-corrected power spectra of $\mathbf{z_{out}}$ and $\mathbf{z_{in}}$ in fast stream, slow stream, the compression region and the rarefaction region at $R=107.1 R_s$ and $R=217.9R_s$. By comparing Figure \ref{fig:spectrum_zinzout_by_regions_RunB} and \ref{fig:spectrum_zinzout_by_regions_RunA}, several differences are observed. First, inside the compression and rarefaction regions (shear regions in Run A), the Els\"asser energies are damped in both runs but in Run B the damping is weaker than Run A. Especially, in Run B the wave energy decays with $k_{x^\prime}$ much slower, indicating that the compression and rarefaction transfer energy from large scales to small scales effectively. Second, in Run B, we also observe an asymmetry between the compression and rarefaction regions: at high-$k_{x^\prime}$ ranges ($k_{x^\prime} R_s \gtrsim 1$), the inward wave energy dominates in the rarefaction region while in the compression region the outward wave energy dominates. Third, inside the fast and slow streams, the evolution of the spectra is different in Run B compared with Run A. At $R=107.1R_s$, the inward waves show $-5/3$ spectra over a substantial range of $k_{x^\prime}$ but the outward waves show spectra steeper than $k_{x^\prime}^{-5/3}$. During the evolution toward $1 \, \mathrm{AU}$, the spectra of $\mathbf{z_{in}}$ steepen while the spectra of $\mathbf{z_{out}}$ develop a Kolmogorov-like inertial range as seen in plot (b) of Figure \ref{fig:spectrum_zinzout_by_regions_RunB}. The span of the inertial range in the fast stream is larger than that in the slow stream.

\begin{figure}[ht!]
\centering
\includegraphics[scale=0.6]{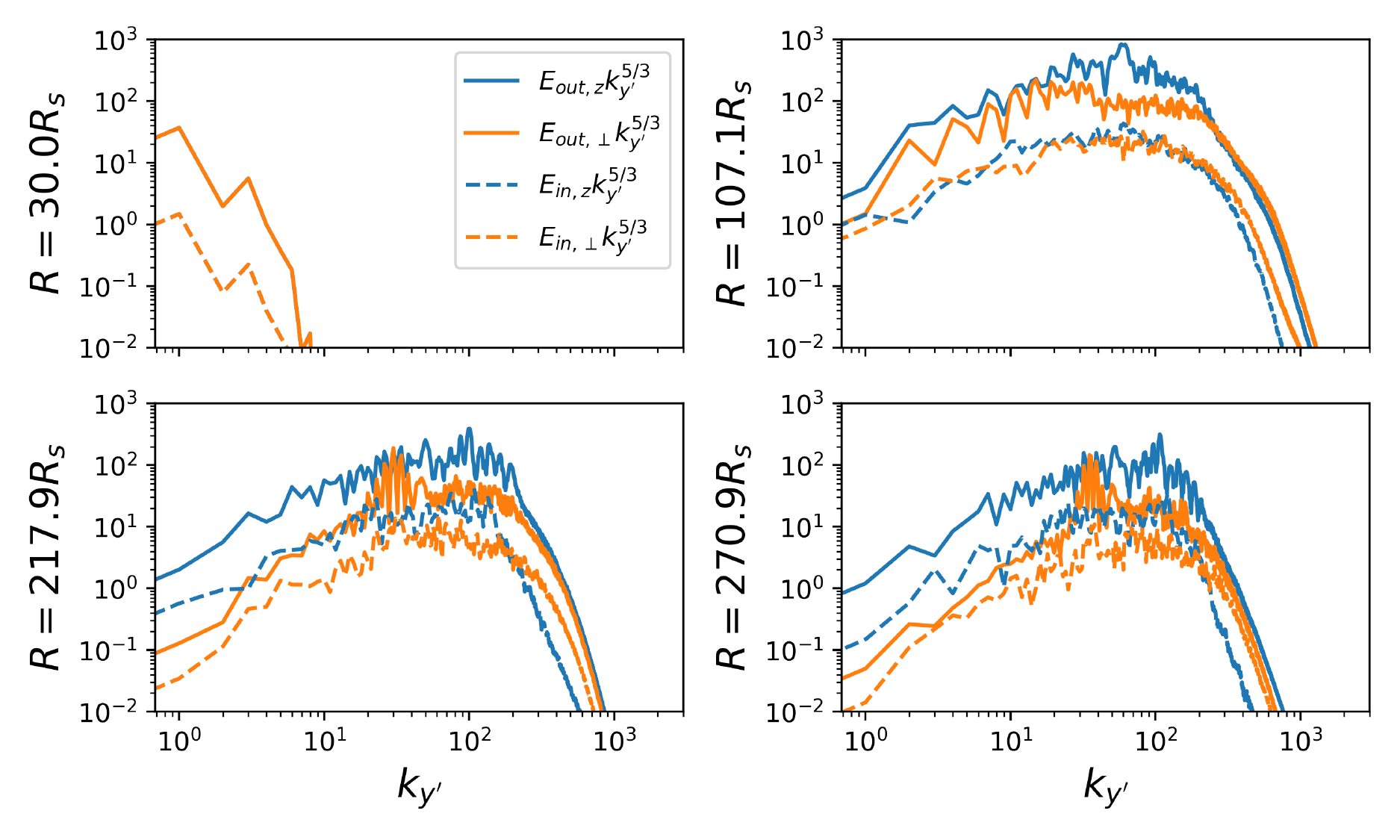}
\caption{Power spectra of the Els\"asser variables calculated along $\hat{e}_{y^\prime}$ and averaged in $x^\prime$ in Run B. From top-left to bottom-right are at $R=30.0R_s, \, 107.1R_s, \, 217.9R_s \, \& \, 270.9R_s$ respectively. Blue and orange solid lines are the $z$-component and in-plane perpendicular-to-$\mathbf{B_0}$-component of the outward Alfv\'en wave. Blue and orange dashed lines are the two components of the inward Alfv\'en wave. The spectra are multiplied by $k_{y^\prime}^{5/3}$. \label{fig:spectrum_zinzout_along_y_RunB}}
\end{figure}

Figure \ref{fig:spectrum_zinzout_along_y_RunB} is the $x^\prime$-averaged $y^\prime$-spectra of $\mathbf{z_{out}}$ and $\mathbf{z_{in}}$ corrected by $k_{y^\prime}^{5/3}$ in Run B. At $R = 107.1 R_s$ the Kolmogorov-type inertial range is well established for both outward and inward waves.  Different from Run A, the shape of the spectra is only slightly changed throughout the simulation in this run.

\subsection{Run C and Run D}

\begin{figure}[ht!]
\centering
\includegraphics[scale=0.7]{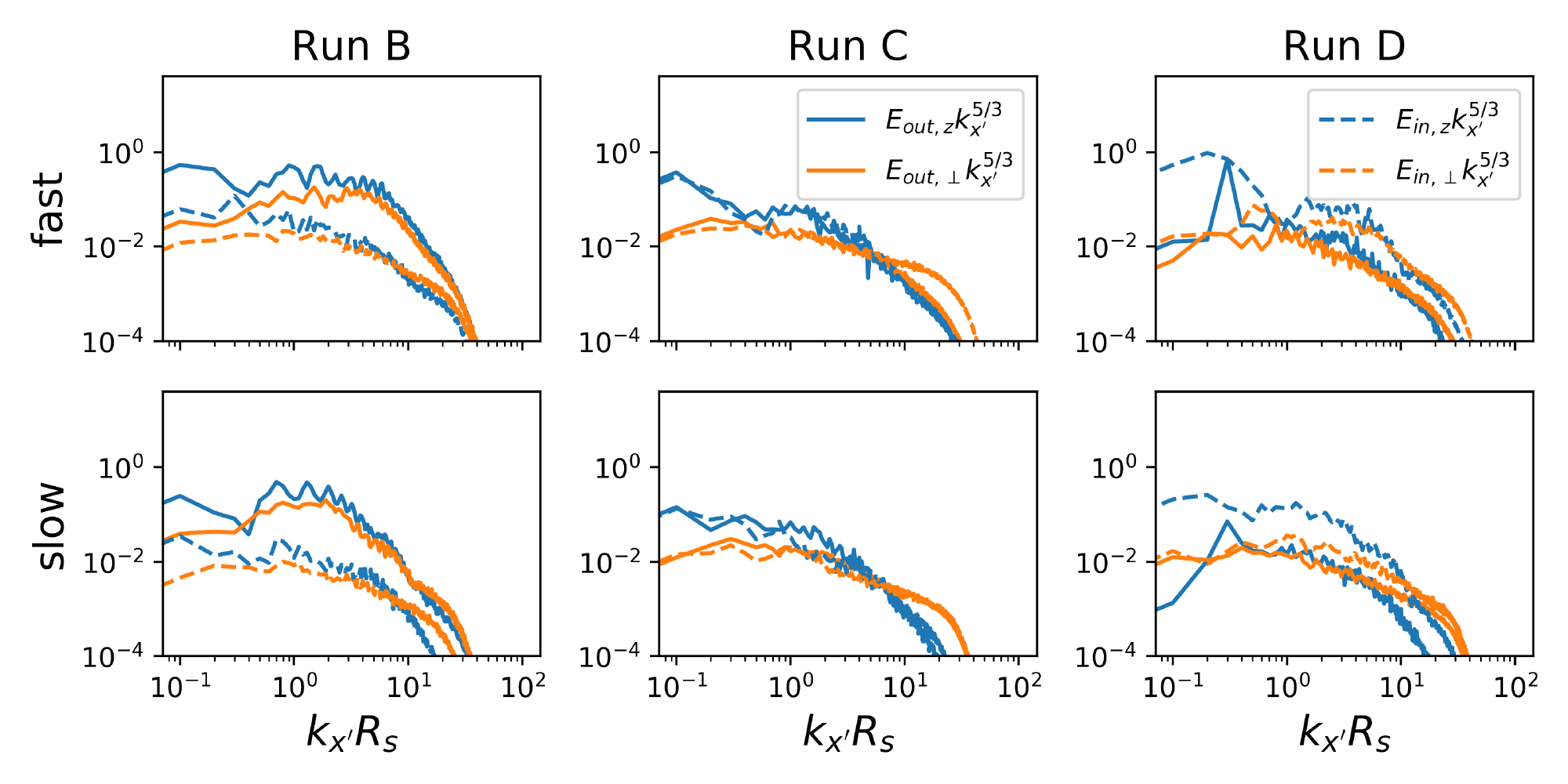}
\caption{Parallel power spectra multiplied by $k_{x^\prime}^{5/3}$ of Els\"asser variables inside the fast (top row) and slow streams (bottom row) at $217.9R_s$ for Run B (left column), C (middle column), and D(right column). Blue and orange solid lines are the $z$-component and in-plane perpendicular-component of the outward Alfv\'en wave. Blue and orange dashed lines are the two components of the inward Alfv\'en wave. \label{fig:compare_parallel_spectrum_RunBCD}}
\end{figure}

Run C and Run D have both corotation and expansion turned on, similar to Run B, but have $r_{io} =1$ and $r_{io} = 5$ respectively. They are carried out to show how the inward and outward waves evolve differently when their amplitudes change. 

\begin{figure}[ht!]
\centering
\includegraphics[scale=0.7]{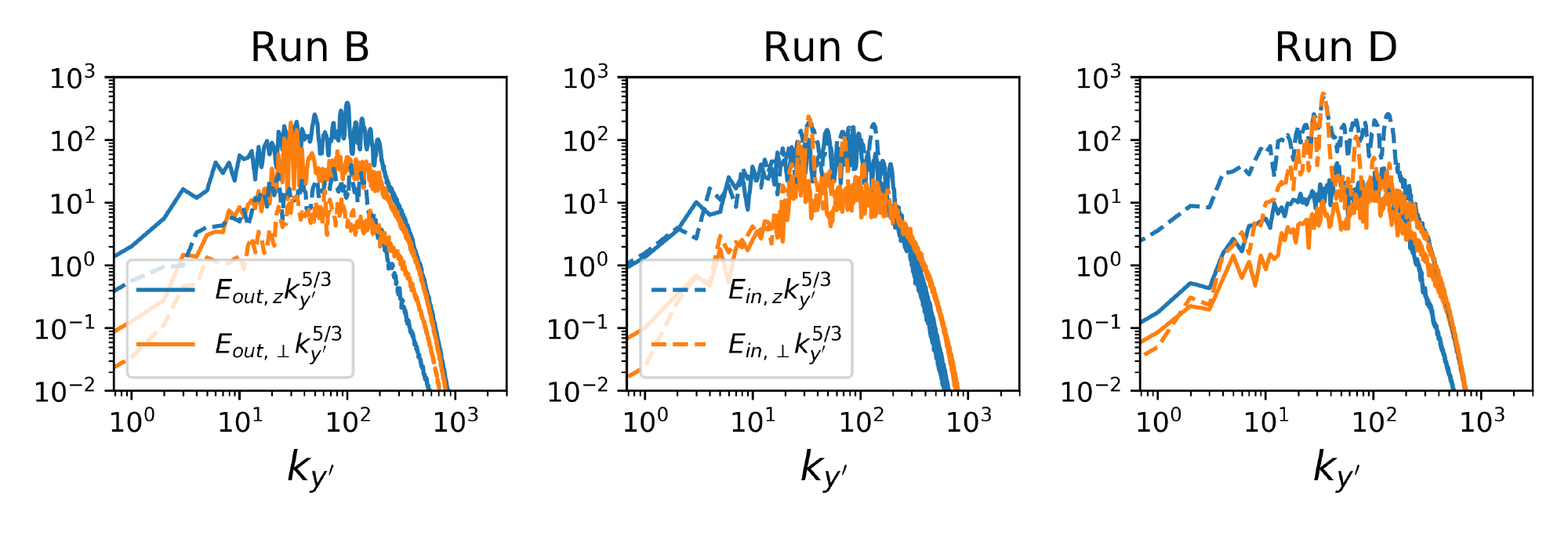}
\caption{Power spectra of Els\"asser variables calculated along $\hat{e}_{y^\prime}$ averaged in $x^\prime$ at $217.9R_s$ for Run B (left), C (middle) and D (right). Blue and orange solid lines are the $z$-component and in-plane perpendicular-component of the outward Alfv\'en wave. Blue and orange dashed lines are the two components of the inward Alfv\'en wave. The spectra are multiplied by $k_{y^\prime}^{5/3}$. \label{fig:compare_perp_spectrum_RunBCD}}
\end{figure}

Figure \ref{fig:compare_parallel_spectrum_RunBCD} compares the $k_{x^\prime}^{5/3}$-corrected parallel power spectra of the Els\"asser variables inside the fast and slow streams at $217.9R_s$ for Run B-D. For Run C and Run D, the spectra inside the slow stream show inertial ranges steeper than $k_{x^\prime}^{-5/3}$. Inside the fast stream, Run D shows a short Kolmogorov-like range at $k_{x^\prime} R_s \sim 1-4$ for $E_{in}$ while Run C shows a shorter one in $E_{out,z}$ and $E_{in,z}$. Note that in Run B, clear Kolmogorov-like inertial ranges are observed in $E_{out}$ spectra inside both fast and slow streams. In other words, in order to get Kolmogorov-like parallel spectra, the outward-dominant initial condition is preferred to the balanced and the inward-dominant ones. However, as shown in Figure \ref{fig:compare_perp_spectrum_RunBCD}, the $k_{y^\prime}$ spectra at $217.9R_s$ are similar for Run B, C and D as clear $k_{y^\prime}^{-5/3}$ inertial ranges are observed in all of the 3 runs. 

\begin{figure}[ht!]
\centering
\includegraphics[scale=0.7]{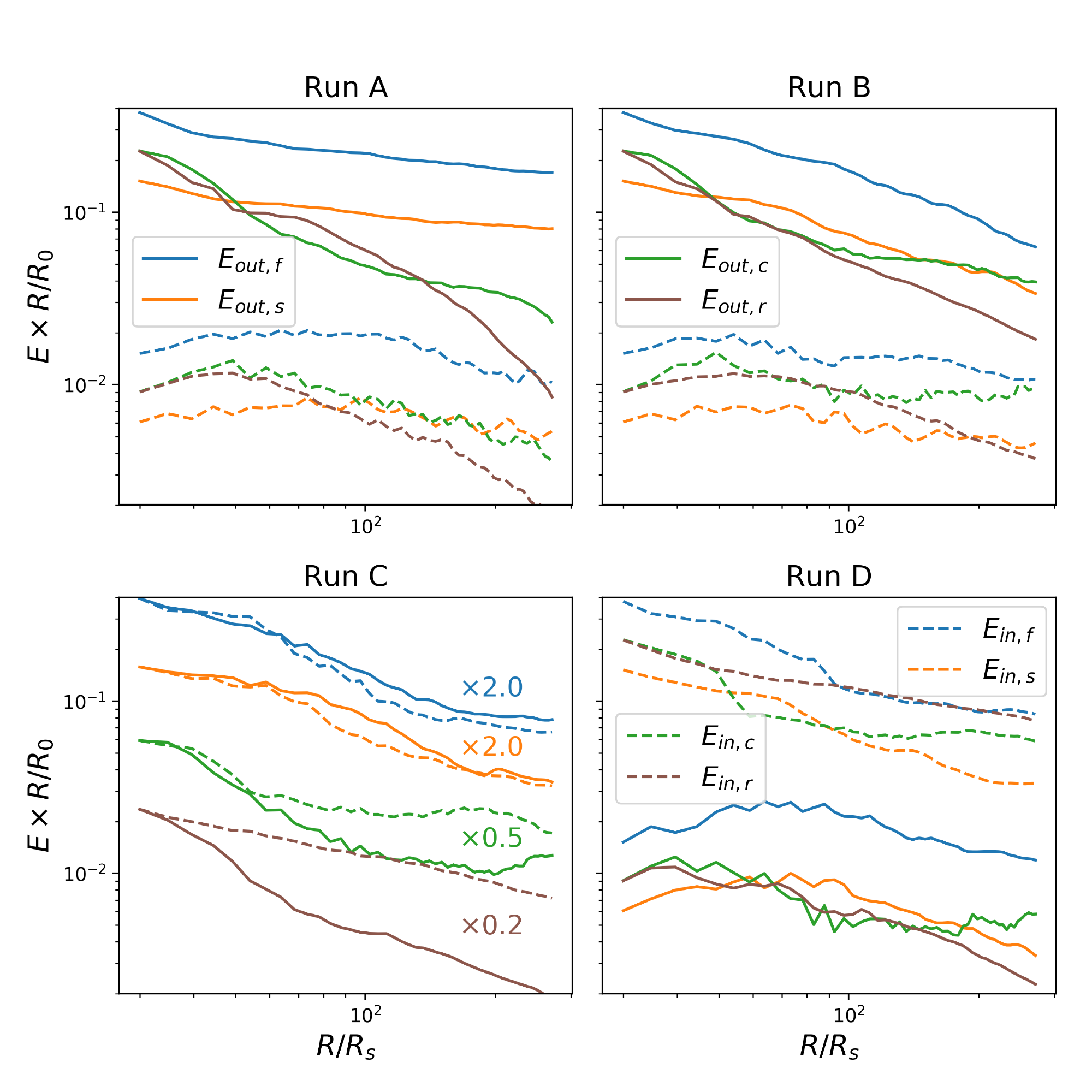}
\caption{Radial evolution of the Els\"asser energies $E_{out}$ (solid curves) and $E_{in}$ (dashed curves) in different regions. The energies are corrected by $R/R_0$. From top-left to bottom-right panels are Run A-D respectively. The plot is in log-log scale. Blue, orange, green and brown represent fast stream (``f''), slow stream (``s''), compression region (``c'') and rarefaction region (``r''). For Run A the compression and the rarefaction regions are the two shear regions around $y^\prime =0.75 L_{y^\prime}$ and $y^\prime =0.25 L_{y^\prime}$ respectively. For Run C we multiply the energies in the four regions by different factors as shown in the plot in order to separate the overlapped curves. \label{fig:radial_evolution_elsasser_energy}}
\end{figure}

We then inspect the radial evolution of $E_{out}$ and $E_{in}$ inside different regions for Run A-D and the results are shown in Figure \ref{fig:radial_evolution_elsasser_energy}. The energies are corrected by $R/R_0$ and the plot is log-log scale. Solid and dashed curves are $E_{out}$ and $E_{in}$ respectively. Colors represent different regions as shown in the legend and described in the caption where the subscripts ``$f$'', ``$s$'', ``$c$'', and ``$r$'' represent fast stream, slow stream, compression region, and rarefaction region respectively. For Run C, we multiply the energies in the four regions by different factors, as shown in the plot, in order to separate the overlapped curves and make the plot more readable. The energies are calculated by averaging $\frac{1}{2}\left| z^\pm \right|^2$ over different regions at each time. We first compare Run A and Run B. These two runs are both outward-dominant but Run A lacks the compression and rarefaction between streams. Compared with Run B, in Run A $E_{out}$ decays much slower inside the fast and slow streams but faster inside the shear regions, i.e. the compression effect speeds up the dissipation of the outward waves in the regions without large velocity gradients but it slows down the dissipation inside the regions with large velocity gradients. The evolution of $E_{in}$ inside the fast and slow streams do not show significant differences between Run A and Run B and approximately follow the $R^{-1}$ WKB prediction. But in the shear regions $E_{in}$, similar to $E_{out}$, decays faster in Run A than Run B. Then we compare Run C with Run B. In Run C the initial condition consists of balanced outward/inward waves instead of outward-dominant waves. By comparing the blue and orange curves in panel Run C, we see that the evolution of $E_{out}$ and $E_{in}$ is very similar to each other inside the fast and slow streams and the decay rates are similar to those of $E_{out}$ in Run B. Inside the rarefaction region, the inward waves decay much slower than the outward waves. Compared to Run B, $E_{in}$ shows a slower decay rate while $E_{out}$ has similar decay rate. In the compression region, both $E_{in}$ and $E_{out}$ show similar evolution as in Run B: a decay followed by a plateau or even an increase. Last, we inspect Run D where the initial condition is inward-dominant wave band, inverse to Run B. Inside the fast stream and the slow stream, $E_{in}$ in Run D evolves similarly with $E_{out}$ in Run B. $E_{out}$ grows at the beginning and then decays with $R$, similar to $E_{in}$ in Run B but its growth and decay are stronger. Consistent with Run C, this result shows that when the wave amplitude is large enough, its radial evolution inside the fast and slow streams is not affected by the direction of the propagation. It is likely that there is some mechanism that generates/depletes small-amplitude waves and it works differently for outward and inward waves. In the compression region, $E_{in}$ in Run D evolves similar with $E_{out}$ in Run B and $E_{out}$ in Run D decreases to a smaller level compared with $E_{in}$ in Run B although both of them reach a plateau beyond $R\approx 10^2 R_s$. In the rarefaction region, $E_{in}$ in Run D has a decay rate similar with that in Run C, i.e. slower than that in Run B. On the other hand, $E_{out}$ in Run B, C and D show very close decay rate beyond $R\approx 10^2 R_s$, indicating that the decay of $E_{out}$ in the rarefaction region is not affected by the wave amplitude significantly.

To summarize the above paragraph, we list the major findings from Figure \ref{fig:radial_evolution_elsasser_energy} below: (1) The radial decrease of the wave amplitude is faster than the WKB prediction when the amplitude is large but gets closer to the WKB prediction when the amplitude is small, especially inside the fast and slow streams where the velocity shear is small. (2) The compression between fast and slow streams speeds up the dissipation of the waves inside the fast and slow streams but slows down the dissipation inside the compression and rarefaction regions. (3) Inside the fast and slow streams, the outward and inward waves do not show significant differences: the radial evolution of their energies are controlled mainly by their amplitudes instead of the propagation directions. (4) In the compression region, the outward wave decays faster than the inward wave but both of them decay slower as the radial distance increases. (5) Inside the rarefaction region, the outward and inward waves show strong asymmetry. The radial decay of the outward wave is in general faster than the inward wave and is not affected by the wave amplitude significantly. The decay of the inward wave energy, on the contrary, is modulated by the wave amplitude: the larger the wave amplitude is, the slower $E_{in}$ decreases with distance.

\section{Conclusion}\label{sec:conclusion}
In this paper, we use the corotating expanding box model (2D version) to simulate the propagation of Alfv\'en waves and turbulence in the solar wind. The large-scale stream interactions, including shear, compression and rarefaction, are evolved self-consistently. The simulation parameters are chosen to be close to the realistic solar wind. We find that the Els\"asser energies are depleted in the strong-shear regions due to  phase-mixing, accompanied by a  decrease of the normalized cross-helicity. This process is greatly enhanced by the compression between fast and slow streams. The generated density fluctuations are overall small ($\delta \rho / \rho \lesssim 0.2$ on average) and there are high-density structures born near the boundaries between the shear regions and the slow streams. The normalized residual energy fluctuates around zero due to the correlation and de-correlation between the outward and inward waves but no net growth or decrease of it is observed, contrary to the solar wind observations which show prevailing excess of magnetic energy \citep[e.g.][]{roberts1987b,grappinetal90,chen2013}. The parallel spectra of the El\"asser variables show Kolmogorov-like inertial ranges only inside the fast and slow streams and when the shear and compression between streams are present. Besides, the outward-dominant waveband is preferred. Otherwise, the parallel spectra are in general steeper than $k^{-5/3}$. On the other hand, the perpendicular, or more precisely the quasi-longitudinal, spectra show Kolmogorov-like inertial ranges in all the runs carried out, no matter whether the compression between streams is present and how the initial waveband is configured. Since the Alfv\'enic fluctuations are in the perpendicular-to-$\mathbf{B_0}$ direction, it is reasonable to expect that the spectra are more developed in this direction (plane). The radial evolution of the El\"asser energies shows significant longitudinal-dependence, symmetry/asymmetry between the outward and inward waves and the wave-amplitude dependence. 

In the present paper we have allowed for the large scale stream structure, but we have not included the corresponding sector structure with heliospheric current sheet. The heliospheric current sheet is known to be embedded inside the slow wind, at least at solar minimum \citep{smith2001}, and the corresponding magnetic shears
might modify the conclusions reached above. Current sheets themselves might evolve dynamically and interact with a turbulence in a non trivial way, as in the region where the magnetic field changes sign, the magnetic field velocity field correlation for outwardly propagating Alfv\'enic fluctuations must also change sign, leaving a region with little correlation and probably a strong magnetic excess in between.  Furthermore, fully three-dimensional simulations need to be carried out for a more realistic solar wind configuration. Third, since the Parker Solar Probe will provide observations at locations from $R \sim 10 R_s$ to $R \lesssim 200 R_s$, it is good to use the data to study the radial evolution of the Alfv\'enic turbulence in the inner heliosphere in the future.

\acknowledgments
This work was supported by the NASA program
LWS, grant NNX15AF34G and by the NSF-DOE
Partnership in Basic Plasma Science and Engineering award n. 1619611. This work used the Extreme Science and Engineering Discovery Environment (XSEDE), which is supported by National Science Foundation grant number ACI-1053575.

%






\end{CJK*}

\begin{thebibliography}{}
 \bibitem[Alazraki \& Couturier (1971)]{alazraki1971} Alazraki, G. \& Couturier, P. 1971, Astron. Astrophys., 13, 380-389
 \bibitem[Bavassano et al.(1982)]{bavassano1982} Bavassano, B., Dobrowolny, M., Fanfoni, G., et al. 1982, 87(A5), 3617-3622
 \bibitem[Belcher(1971)]{belcher1971} Belcher, J. W. 1971, ApJ, 168, 509-524
 \bibitem[Belcher \& Davis(1971)]{belcherdavis1971} Belcher, J. W. \& Davis Jr., L. 1971, J. Geophys. Res., 76(16), 3534-3563
 \bibitem[Bruno \& Bavassano(1991)]{bruno1991} Bruno, R. \& Bavassano, B. 1991, J. Geophys. Res., 96, A5, 7841-7851
 \bibitem[Chen et al.(2013)]{chen2013} Chen, C. H. K., Bale, S. D., Salem, C. S., et al. 2013, ApJ, 770:125
 \bibitem[Coleman(1968)]{coleman1968} Coleman Jr., P. J. 1968, ApJ, 153, 371
 \bibitem[Dobrowolny(1980)]{dobrowolny1980} Dobrowolny, W., Mangeney, A. \& Veltri, P. L. 1980, Solar and Interplanetary Dynamics, 143-146
 \bibitem[Goldstein et al.(1989)]{goldstein1989} Goldstein, M. L., Roberts, D. A. \& Matthaeus, W. H. 1989, Solar System Plasma Physics, 54, 113
 \bibitem[Grappin \& Mangeney(1990)]{grappinandmangeney1990} Grappin, R. \& Mangeney, A. 1990, J. Geophys. Res., 95(A6), 8197-8209
 \bibitem[Grappin \& Velli(1996)]{grappin1996} Grappin, R. \& Velli, M. 1996, J. Geophys. Res., 101, 425-444
 \bibitem[Grappin et al.(1993)]{grappin1993} Grappin, R., Velli, M. \& Mangeney, A. 1993, Phys. Rev. Lett., 70, 2190
 \bibitem[Grappin et al.(1990)]{grappinetal90}Grappin, R.,  Mangeney, A. and Marsch, E., 1990
 \bibitem[Heinemann \& Olbert(1980)]{heinemann1980} Heinemann, M. \& Olbert, S. 1980, J. Geophys. Res., 85(A3), 1311-1327
 \bibitem[Hollweg(1974)]{hollweg1974} Hollweg, J. V. 1974, J. Geophys. Res., 79, 10
 \bibitem[Iroshnikov(1964)]{iroshnikov1964} Iroshnikov, P. S. 1964, Soviet Astronomy, 7, 566
 \bibitem[Kraichnan(1965)]{kraichnan1965} Kraichnan, R. H. 1965, PhFl, 8, 1385
 \bibitem[Lele(1992)]{lele1992} Lele, S. K. 1992, J. Computational Physics, 103, 16-42
 \bibitem[Marsch \& Tu(1990)]{marsch1990} Marsch, E. \& Tu, C.-Y. 1990, J. Geophys. Res., 95(A6), 8197-8209
 \bibitem[Roberts et al.(1987a)]{roberts1987a} Roberts, D. A., Goldstein, M. L., Klein, L. W., et al. 1987a, J. Geophys. Res.: Space Physics, 92(A11), 12023-12035
 \bibitem[Roberts et al.(1992)]{roberts1992} Roberts, D. A., Goldstein, M. L., Matthaeus, W. H., et al. 1992, J. Geophys. Res.: Space Physics, 97(A11), 17115-17130
 \bibitem[Roberts et al.(1987b)]{roberts1987b} Roberts, D. A., Klein, L. W., Goldstein, M. L., et al. 1987b, J. Geophys. Res.: Space Physics, 92(A10), 11021-11040
 \bibitem[Smith(2001)]{smith2001} Smith, E. J. 2001, J. Geophys. Res. Space Physics, 106, A8, 15819-15831
 \bibitem[Tenerani \& Velli(2017)]{tenerani2017} Tenerani, A. \& Velli, M. 2017, ApJ, 843(1), 26
 \bibitem[Tu et al.(1984)]{tuetal1984} Tu, C. Y., Pu, Z. Y. \& Wei, F. S. 1984, J. Geophys. Res.: Space Physics, 89(A11), 9695-9702
 \bibitem[Velli(1993)]{velli1993} Velli, M. 1993, Astron. Astrophys., 270, 304-314
 \bibitem[Velli et al.(1991)]{velli1991} Velli, M., Grappin, R. \& Mangeney, A. 1991, Geophysical and Astrophysical Fluid Dynamics, 62: 1, 101-121
 \bibitem[Zank et al.(1996)]{zank1996} Zank, G. P., Matthaeus, W. H. \& Smith, C. W. 1996, J. Geophys. Res., 101(A8), 17093-17107
 \bibitem[Zank et al.(2012)]{zank2012} Zank, G. P., Dosch, A., Hunana, P., et al. 2012, ApJ, 745, 35
 \bibitem[Zhou \& Matthaeus(1990)]{zhou1990} Zhou, Y. \& Matthaeus, W. H. 1990, J. Geophys. Res., 95(A7), 10291-10311
\end{thebibliography}
\end{document}